\documentclass[12pt]{iopart}

\newcommand{\be}{\begin{equation}}
\newcommand{\ee}{\end{equation}}
\newcommand{\bea}{\begin{eqnarray}}
\newcommand{\eea}{\end{eqnarray}}
\usepackage{iopams}  
\usepackage{graphicx}

\begin{document}

\title{Topological Boundary States in $1$D: An Effective Fabry-Perot Model}

\author{E. Levy$^{1}$, E. Akkermans$^{2,*}$}

\address{$^1$Rafael Ltd., P.O. Box 2250, Haifa 32100, Israel}
\address{$^2$Department of Physics, Technion Israel Institute of Technology, Haifa 32000, Israel}
\address{$^*$Corresponding author}

\ead{eric@physics.technion.ac.il}
\vspace{10pt}
\begin{indented}
\item[]October 2016
\end{indented}

\begin{abstract}
We present a general and useful method to predict the existence, frequency, and spatial properties of gap states in photonic (and other) structures with a gapped spectrum. This method is established using the scattering approach. It offers a viewpoint based on a geometrical Fabry-Perot model. We demonstrate the capabilities of this model by predicting the behaviour of topological edge states in quasi-periodic structures. A proposition to use this model in Casimir physics is presented.
\end{abstract}

%
%
%
%
%


\section{Introduction}
One-dimensional multilayer photonic structures have proven to be important in fundamental and applied physics as being simple and accessible systems which display an equivalence to solid state systems. Periodic dielectric structures are, to some extent, a photonic counterpart of a solid state crystal whose spectrum contains both transmission bands and gaps \cite{yablonovitch1987}. Random dielectric structures have been used to study  Anderson localization of light \cite{anderson1958, genack1991, storzer2006, evers2008, maret2013, bellando2014, maret2016, skipetrov2016}, and quasiperiodic structures have been shown to be a photonic analogue of a solid state quasicrystal with a fractal spectrum containing infinitely many gaps \cite{kohmoto1987,gellermann1994,akkermans2013evgeni,akkermans2013,Tanese2014}.
Gap states, which may occur in any gapped spectrum are very useful due to their relatively high spectral isolation and spatial localization. Several methods exist to generate such states, the most familiar being defect states in periodic systems, similar to the insertion of a dopant into a semiconductor crystal \cite{joannopoulos2011}. In the $1$D case, a local defect inserted into a periodic chain gives rise to a defect state in the gap, spatially localized around the defect, with perfect transmission inside the band-gap.
Also for the periodic case and under certain choice of boundary conditions, a perfect photonic crystal may also give rise to surface (or edge) gap states, where the electromagnetic field is localized at the boundary \cite{joannopoulos2011} (In solid state electronic systems these states are termed Tamm and Shockley states \cite{tamm1932,shockley1939}).
A third scheme known to generate gap states, is a heterostructure composed of the concatenation of two periodic chains of different periods or different unit cells, where the gap states are localized at the interface \cite{goto2008}.
In all of the above cases, the existence of gap states is attributed to the breaking of the crystalline discrete translational invariance by the structural defect or boundary. For photonic crystals, the energy spectrum cannot contain extended states with purely real wave vectors within the gaps. But evanescent states with complex wave vectors are legitimate and allowed solutions of the eigenvalue problem within the gaps. Therefore, any symmetry breaking feature such as a structural defect or a boundary may generate evanescent gap states localized around the defect or at the boundary respectively. When applying these considerations to quasicrystals, this understanding deserves a closer look and further generalization.
Although less known than the periodic case, gap states may be easily induced in spectral gaps of quasiperiodic structures. For instance, in the case of the Fibonacci photonic quasiperiodic chain, gap states of all kinds have been observed: defect \cite{Abdel-Rahman2009}, boundary \cite{Zijlstra1999,pang2010} and interface \cite{Hiltunen2007,Zhukovsky2010} gap states. The origin of such gap states in structures lacking a definite spatial symmetry is not that obvious. As a result of this lack of clarity, these gap states have been given various names, such as “defect states”, “perfect transmission resonances”, and various explanations have been given for their origin (see e.g. \cite{Kalozoumis2013}).

In this note, we start from the fact that along with their vanishingly small density-of-states (hereafter DOS) and transmittance values, spectral gaps are also characterized by high reflectance. We argue that the notion of gap states generated by boundary conditions bears a Fabry-Perot like meaning, and in that sense, many seemingly different schemes used to produce gap states are actually of the same origin. Using the scattering formalism \cite{panbookchapter}, we derive a generalised effective Fabry-Perot condition to analyze gap states. We then demonstrate this very general geometric viewpoint in analyzing edge and interface states which have been recently used to characterise topological properties of quasiperiodic chains \cite{levy2015}. In this test case, the Fabry-Perot description allows to obtain the spectral locations of gap states and also to fully characterize their topological content mapped onto a cavity effective length, a result not easily anticipated. 
 
This note is organised as follows. Section 2 describes the scattering approach, section 3 presents the effective Fabry-Perot model and section 4 its implementation to the characterisation of topological boundary states for a 1D Fibonacci quasicrystal. Finally, section 5 concludes and discusses our results. 


\section{Scattering analysis of $1$D structures}
\subsection{Scattering phases - Total phase shift and spectral properties}
We now introduce the scattering approach which offers a framework to study spectral properties of  any structures including those without obvious spatial symmetry, such as quasiperiodic structures \cite{panbookchapter}. We recall that any quantum or wave system with a potential defined w.r.t a free part can be probed using the scattering of waves with wave vector $k$. For the sake of self coherence and simplicity of notations, we consider the scattering of electromagnetic waves by a dielectric medium with a spatially modulated refractive index. 

\begin{figure}
\includegraphics[viewport=-155bp 16bp 1088bp 531bp, clip,width=1\columnwidth]{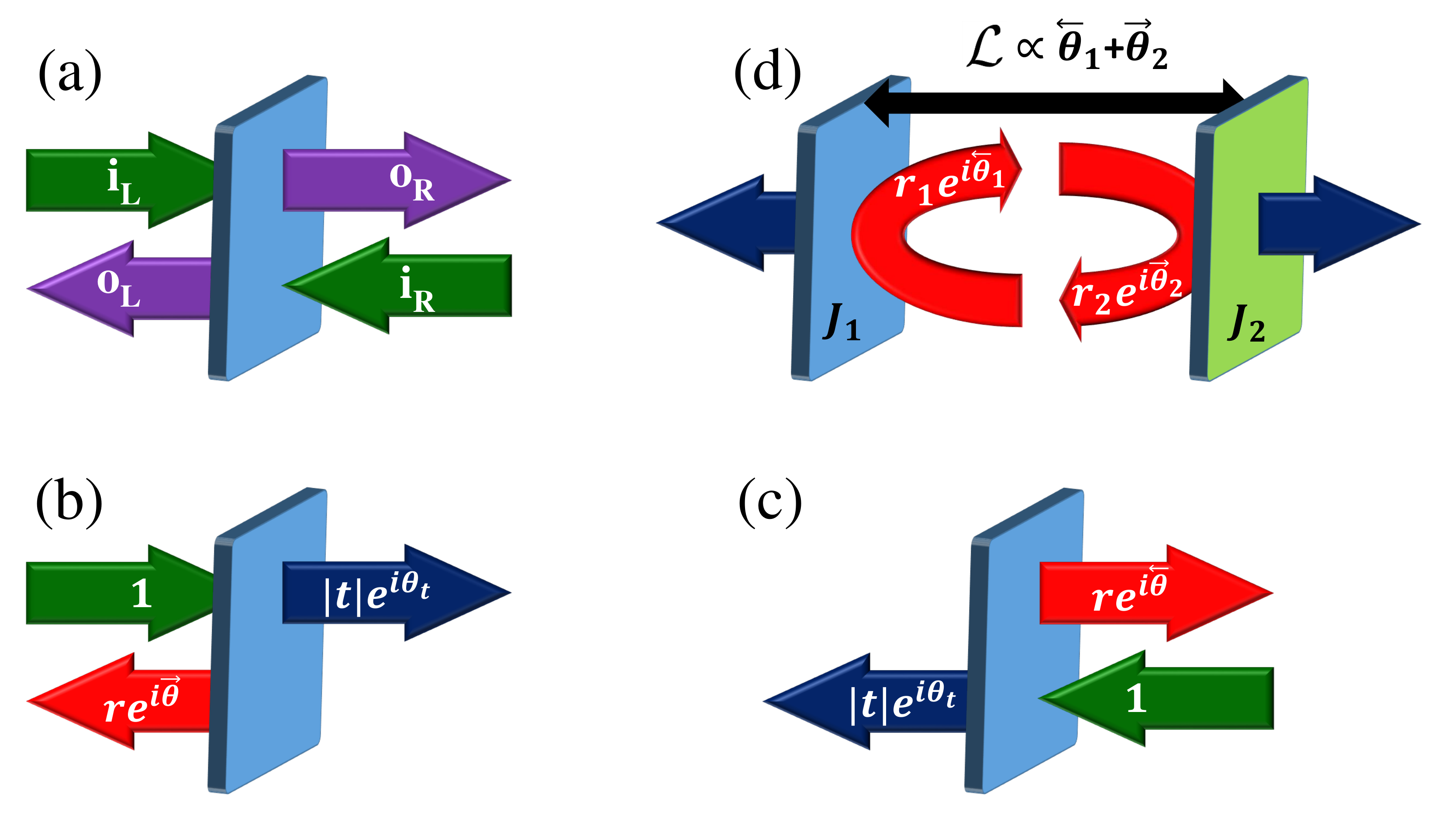}

\caption{(color online) The scattering problem. (a) A sketch for the notations of incoming
and outgoing waves. (b)-(c) Notations for the amplitude of the two possible transmission experiments:
incoming waves from the right or from the left. (d) Notations for the virtual cavity in the interface between two general structures $J_1$ and $J_2$.\label{FPsketch}}
\end{figure}

The scattering matrix $\mathcal{S}$ of a $1$D structure is defined by
\bea
 \left(
\begin{array}{l}
o_L  \\
o_R
\end{array}
 \right)
 =  \left(
\begin{array}{ll}
\overrightarrow{r} \, & t \,  \\
t \,  & \overleftarrow{r} \,
\end{array}
 \right)  \left(
\begin{array}{l}
i_L  \\
i_R
\end{array}
 \right)
 \equiv \mathcal{S}\,
 \left(
\begin{array}{l}
i_L  \\
i_R
\end{array}
 \right) \, ,
\label{smatrix_new}
\eea
where $\left[i,o\right]$ stand for incoming and outgoing plane waves, $\left[R,L\right]$ stand for right-hand and left-hand side of the structure, $\left[\overrightarrow{r},\overleftarrow{r}\right]$ stand for the reflected wave amplitudes corresponding to incoming waves from the left and from the right respectively, and $t$ is the transmitted amplitude (see Fig. \ref{FPsketch}a). The scattering matrix $\mathcal{S}$ is unitary, so that it is diagonalizable under the form 
\be
\left(
\begin{array}{ll}
 e^{i \Phi_1 (k)} \, & 0 \,  \\
0 \,  &  e^{i \Phi_2 (k) } \,
\end{array}
 \right),
\label{smatrix_diag}
\ee
and therefore may be fully described by means of two independent scattering phases. One convenient choice of such phases is
\be
\begin{array}{ll}
\delta (k)\! \equiv \! \left( \Phi_1  + \Phi_2  \right) /2\,  \\
\Lambda (k)\equiv\!(\Phi_2  - \Phi_1  )/2 \,
\end{array},
\label{scatt_phases}
\ee
The scattering phase $\delta(k)$, known as the total phase shift, allows to obtain a simple and very useful relation to $\rho(k)$, the DOS \cite{panbookchapter}, sometimes known as the Krein-Schwinger formula, namely  
 \be
\rho (k) - \rho_0 (k) = {1 \over 2 \pi}\, \mbox{Im} \,\frac{\partial}{\partial k} \ln \mbox{det}\, \mathcal{S}(k) = {1 \over \pi} {d \delta (k) \over d k} \, , 
\label{DOS}
\ee
where $\rho_0(k)$ is the DOS of the free system, i.e. without  the scattering structure.
Defining the transmission and reflection complex amplitudes by 
\be
\begin{array}{ll}
 t\!\equiv \!|t|e^{i \, \theta_t} \,  \\
\overrightarrow{r}\!\equiv \!r e^{i \overrightarrow{\theta}}  \,  \\
\overleftarrow{r} \!\equiv\! r e^{i \overleftarrow{\theta}} \,
\end{array}
\label{scatt_amps}
\ee
again with the same arrow convention indicating incoming waves from left or right, and using the unitarity condition, $\overrightarrow{r}\! \, ^* \, t+\overleftarrow{r} \,  \!t^*=0$, we obtain the additional expressions, 
\be
\begin{array}{ll}
\mbox{det}\, \mathcal{S} \!=\! e^{2 i \delta }\! = \!- t / t^*=\overleftarrow{r} \!/\overrightarrow{r}\! \, ^*\,  \\
\delta(k)\!=\!\theta_t (k) \!+\!\pi/2=  {1 \over 2 } \left( \overrightarrow{\theta} + \overleftarrow{\theta} \right) \, .
\end{array}
\label{scatt_total_ps}
\ee
Note that the relations (\ref{scatt_total_ps}) and the whole scattering approach apply for finite structures (where $\left[t, \overrightarrow{r},\overleftarrow{r}\right]$ are well defined), and are equally correct in the limit $r^2=1-|t|^2 \rightarrow 1$. Also notice that the notations  $\overrightarrow{r}$ and $\overleftarrow{r}$ in (\ref{smatrix_new}) and (\ref{scatt_total_ps}) represent the two possible transmission experiments (see Figs. \ref{FPsketch}b and \ref{FPsketch}c) which are identical except for the phases of the reflected amplitudes. Therefore, the total phase shift may be expressed using either the transmitted phase shift or the sum of the two possible reflected phase shifts.
The second scattering phase $\Lambda(k)$ in (\ref{scatt_phases}), complementary to $\delta(k)$, carries additional information regarding the structure, unavailable through $\delta(k)$. Here we wish to promote the use of a related phase, termed the chiral scattering phase \cite{levy2015} and defined by $\overleftarrow{r} \!\equiv \! \overrightarrow{r} \, e^{i \alpha}$, or equivalently by
\be
\alpha \!\equiv \!\overleftarrow{\theta} \!-\! \overrightarrow{\theta}.
\label{scatt_chiral_phase}
\ee 
This phase, which monitors the difference between the two possible transmission experiments in Figs. \ref{FPsketch}b and \ref{FPsketch}c, is equivalent to the phase $2\Lambda(k)$ in the limit $r^2=1-|t|^2 \rightarrow 1$ as will be shown in the next subsection. The chiral phase has been shown to be related to the topological nature of the spectrum \cite{levy2015}, as we shall see later on. 
Relations  (\ref{scatt_total_ps}) and (\ref{scatt_chiral_phase}) are the starting point in deriving the generalised effective Fabry-Perot model.
\subsection{The chiral scattering phase $\alpha$}\label{whatisalpha}
The scattering matrix defined in (\ref{smatrix_new}) may be rewritten using (\ref{scatt_amps})-(\ref{scatt_chiral_phase}) as
\begin{equation}
\mathcal{S}=e^{i\delta}\left(\begin{array}{cc}
re^{-\frac{i\alpha}{2}} & i\sqrt{1-r^{2}}\\
i\sqrt{1-r^{2}} & re^{\frac{i\alpha}{2}}
\end{array}\right)\equiv e^{i\delta}\tilde{\mathcal{S}}.
\end{equation}
The phases $\Phi_1$ and $\Phi_2$  in (\ref{smatrix_diag}) may now be calculated through the diagonalization of $\tilde{\mathcal{S}}$ so that
\begin{equation}
{P}^{-1}\tilde{\mathcal{S}}{P}=\left(\begin{array}{cc}
e^{i\left(\Phi_{1}-\delta\right)} & 0\\
0 & e^{i\left(\Phi_{2}-\delta\right)}
\end{array}\right).
\end{equation}
The $2$ eigenvalues of $\tilde{\mathcal{S}}$ are the solutions $\lambda_{1,2}=e^{i\left(\Phi_{1,2}-\delta\right)}$
of
\begin{equation}
\lambda^{2}-2r\lambda\cos\frac{\alpha}{2}+1=0,
\end{equation}
namely
\begin{equation}
\lambda_{1,2}=r\cos\frac{\alpha}{2}\pm i\sqrt{1-r^{2}\cos^{2}\frac{\alpha}{2}} \, \, .
\end{equation}
This implies that
\begin{equation}
\cos\left[\pm\left(\Phi_{1,2}-\delta\right)\right]=r\cos\frac{\alpha}{2},
\end{equation}
or equivalently,
\begin{equation}
\Phi_{1,2}=\delta\pm\arccos\left(r\cos\frac{\alpha}{2}\right) \, ,
\end{equation}
which leads to the useful representation of (\ref{scatt_phases})
\begin{eqnarray}
\Phi_{2}+\Phi_{1} & = & 2\delta\;\;\mbox{for any \ensuremath{r}}
\label{independent_phases1}\\
\Phi_{2}-\Phi_{1} & = & 2\arccos\left(r\cos\frac{\alpha}{2}\right) = 2 \Lambda \, \, .
\label{independent_phases2}
\end{eqnarray}
Expression (\ref{independent_phases2}) gives a precise relation between the phases $\Lambda$ and $\alpha$ and the reflectance. We can investigate this relation by writing
\begin{eqnarray}
\cos\frac{\Phi_{2}-\Phi_{1}}{2} & = & r\cos\frac{\alpha}{2}.
\label{alfa_lambda}
\end{eqnarray}
From (\ref{alfa_lambda}) we can see that $\Phi_{2}-\Phi_{1}=\alpha$ for $r\rightarrow1$, but also for every value of $r$ when $\alpha=\pi$. This means that for any nonzero value of $r$, the scattering phase $\Phi_{2}-\Phi_{1}$ follows the behavior of $\alpha$, but attenuated by the factor $0<r<1$ for most part of the winding of $\alpha$. This behavior is the reason why we advocate the use of the scattering phase shift $\alpha$, which is a key quantity in the generalised effective Fabry-Perot model.


\section{The effective Fabry-Perot model}\label{FP}
\subsection{Fabry-Perot interference condition and the winding of a phase}
We begin with a qualitative argument. Any structure with a gapped spectrum is equivalent to a well defined frequency-dependent mirror. Each of these multilayered mirrors (with distributed reflection) is equivalent to a standard mirror with a frequency dependent reflectance, and a frequency dependent phase shift upon reflection. We will show that a cavity delimited by such multilayered mirrors, is equivalent to a Fabry-Perot cavity with standard (phase conserving) mirrors and an effective (i.e. often non geometric) cavity length.

In the usual Fabry-Perot description, a cavity is defined by two mirrors separated by a geometrical length $L$. Such a cavity contains a discrete set of resonant wavelengths, $\lambda_{m}$, obtained from the standard  resonance condition 
\be
2L/\lambda_{m}= m,  \, \, \, m\!\in\!\mathbb{Z}.
\label{FP_res}
\ee
Equivalently, this resonance condition can be written in terms of the winding of a new frequency dependent round-trip accumulated phase, $\theta^{(0)} _{cav}$, termed the cavity phase and defined by
\be
\theta^{(0)} _{cav}\left(k,L\right)\equiv\frac{4\pi L}{\lambda\left(k\right)},
\ee
where $\lambda(k)\!=\!2\pi/k$ is the wavelength. Under this definition, the resonance condition in (\ref{FP_res}) becomes
\be
\theta^{(0)} _{cav}(k_{m})=2\pi m$, \; $m\!\in\!\mathbb{Z}.
\label{FP_res2}
\ee 
We now wish to generalise these usual interference conditions and phase winding so as to include the multilayered mirrors $J_1$ and $J_2$ on the left hand and right hand side of the cavity respectively (see Fig. \ref{FPsketch}d). The mirrors $J_1$ and $J_2$  lend a non zero phase upon reflection, $\overleftarrow{\theta}_{1}\left(k\right)$ and $\overrightarrow{\theta}_{2}\left(k\right)$ respectively, allowing to write a new and generalised cavity phase as
\be
\theta_{cav}\left(k,L,\overleftarrow{\theta}_{1} ,\overrightarrow{\theta}_{2} \right)  \equiv \, \, \theta^{(0)} _{cav}+\!\overleftarrow{\theta}_{1} +\!\overrightarrow{\theta}_{2} \!.
\label{thetacav}
 \ee
for which the resonance condition still applies by the generalisation of (\ref{FP_res2}),
\be
\theta _{cav}(k_{m})=2\pi m$, \; $m\!\in\!\mathbb{Z} 
\label{FP_res2gen}
\ee 
Since this resonance condition holds also in the absence of a geometrical cavity namely for two adjacent multilayer structures in the absence of separation, i.e. for which $ L =0$ and $\theta^{(0)} _{cav}=0$, it is thus possible to define a virtual cavity which still admits  resonant states. This winding condition can also be formulated as a Fabry-Perot condition (\ref{FP_res}) but where the geometric length $L$ is now replaced by a frequency-dependent effective cavity length, ${\cal L}\left(k\right)$, built out of the sum of the geometrical and the virtual cavity lengths, namely
\be
{\cal L}\left(k,L,\overleftarrow{\theta}_{1} ,\overrightarrow{\theta}_{2} \right) \equiv \frac{\lambda(k)}{4\pi}\,\theta_{cav} .
\label{effect_length}
\ee
Resonant states occur at gap frequencies satisfying the usual Fabry-Perot condition, this time for the effective length, namely
\be
2{\cal L}(k_{m})/\lambda(k_{m})\!=\! m, \, \, \, m\!\in\!\mathbb{Z}.
\label{FP_res3}
\ee
We emphasize that although both (\ref{FP_res2}) and (\ref{FP_res3}) may hold for many values of $k$, only a discrete set $\left\{ k_{m}\right\}$ of values within the spectral gaps (where the reflectance values are sufficiently high) will support gap states. The figure of merit used to predict the existence of gap states is the mutual reflectance $\mathcal{R}$, to be defined later on.
Additionally, it is possible to show that the $\left\{ k_{m}\right\}$ values which fall within transmission bands will result in so called perfect transmission resonances \cite{Zhukovsky2010,Kalozoumis2013}, but this lies outside the scope of this note.


\subsection{Fabry-Perot interference condition from scattering theory} 

The idea of a frequency dependent cavity length can lead, in the case of a nontrivial $k$-dependence in ${\cal L}(k)$, to some counterintuitive results such as a non-equal spacing of resonant modes in frequency (unlike the traditional Fabry-Perot situation), and even to the manifestation of a single Fabry-Perot mode spread into more than one resonant frequencies. However, this idea of a frequency dependent cavity length is necessary even in trivial cases, as shown in \ref{WDCL}.

\begin{figure}
\includegraphics[viewport=-155bp 16bp 988bp 481bp, clip,width=1\columnwidth]{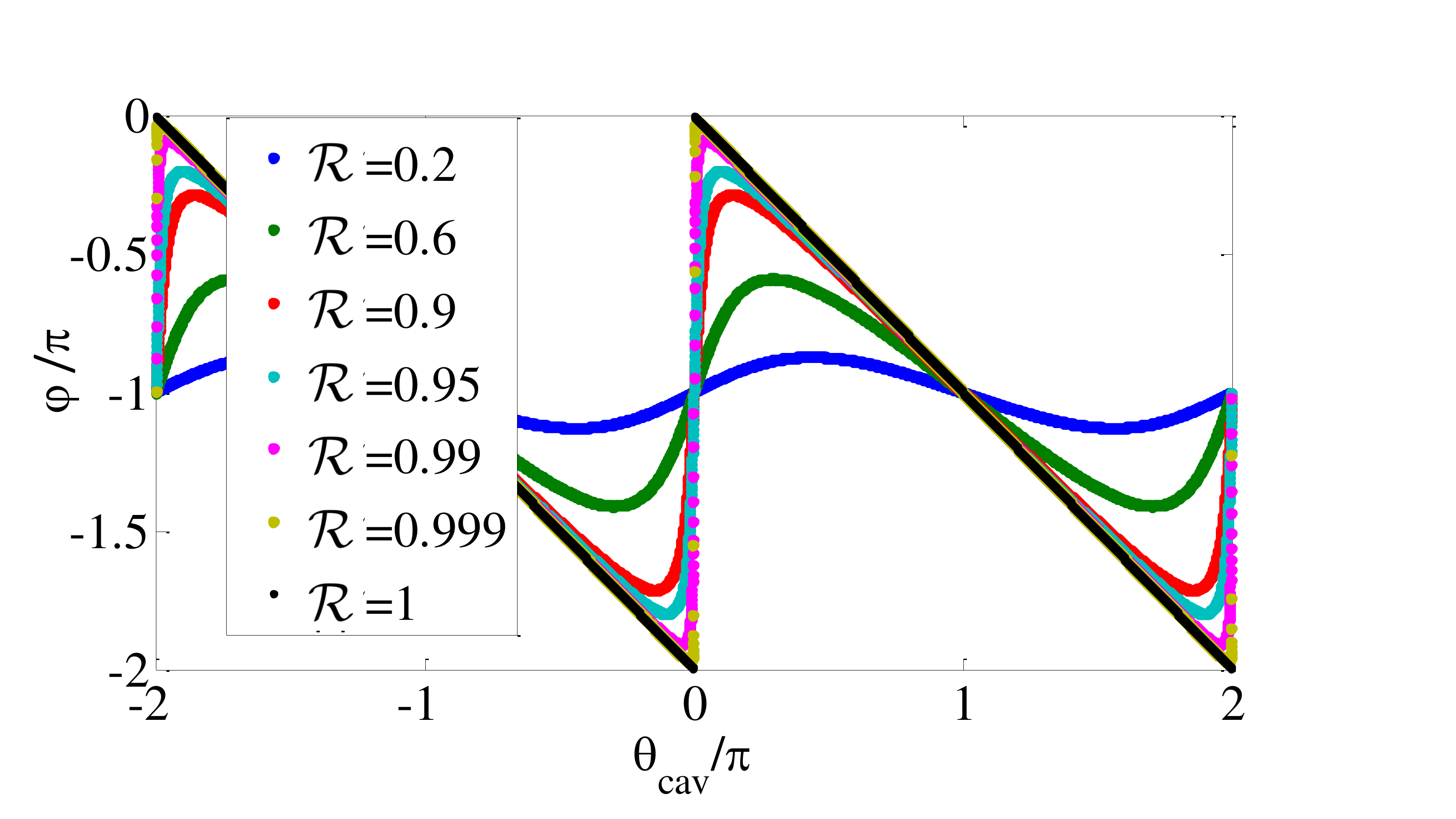}
\caption{(color online) $\varphi(\theta_{cav})$ for various values of the mutual reflectance $\mathcal{R}$ as calculated from (\ref{trigo2}). The relation $\varphi=-\theta_{cav}$ holds only for perfect mutual reflectance $\mathcal{\mathcal{{{R}}}}=1$.\label{varphi_tc}}
\end{figure}

The results (\ref{FP_res2gen}) and (\ref{FP_res3}) can be obtained in a more quantitative way for the case $L=0$, namely for the heterostructure $\left[J_{1}J_{2}\right]$ made of the concatenation of  the two sub-structures $J_{1}$ and $J_{2}$ i.e. without  geometric cavity. There exists a relation between the scattering matrix $\mathcal{S}_{12}\left(k\right)$ of the heterostructure and the scattering matrices $\mathcal{S}_{1}\left(k\right),\,\mathcal{S}_{2}\left(k\right)$ of the respective substructures $J_{1}$ and $J_{2}$. This relation is obtained starting from the definitions 
\be
\mathcal{S}_{1(2)}\equiv \left(\begin{array}{cc}
\overrightarrow{r}_{1(2)} & t_{1(2)}\\
t_{1(2)} & \overleftarrow{r}_{1(2)}
\end{array}\right)\,;\,\mathcal{S}_{12}\equiv \left(\begin{array}{cc}
\overrightarrow{R} & T\\
T & \overleftarrow{R}
\end{array}\right).
\ee 
Then, using the algebraically equivalent transfer matrix \cite{panbookchapter} defined by
\be
\left(
\begin{array}{c}
o_{R}\\
i_{R}
\end{array}
\right)
 \equiv M\left
(\begin{array}{c}
i_{L}\\
o_{L}
\end{array}
\right),
\ee
such that
\be
M_{1(2)}=\left(\begin{array}{cc}
{1 / t_{1(2)}^{*}} & {\overleftarrow{r}_{1(2)} / t_{1(2)}}\\
-\overrightarrow{r}_{1(2)} / t_{1(2)} & 1 / t_{1(2)}
\end{array}\right)\,;\, 
\ee
and
\be
M_{12}=M_{1}\cdot M_{2}\equiv\left(\begin{array}{cc}
1 / T^{*}  & {\overleftarrow{R}  / T} \\
-\overrightarrow{R} / T & 1 / T
\end{array}\right) \, .
 \ee
The total phase shifts $\mathcal{\delta}_{1},\,\mathcal{\delta}_{2},\,\mathcal{\delta}_{12}$ of the respective scattering matrices $\mathcal{S}_{1},\,\mathcal{S}_{2},\,\mathcal{S}_{12}$ are obtained from
\be
e^{2i\delta_{1(2)}}=det \, \mathcal{S}_{1(2)}=-\frac{t_{\mathbf{1(2)}}}{t_{1(2)}^{*}}=\frac{\overleftarrow{r}_{1(2)}}{\overrightarrow{r}_{1(2)}^{*}}\;;\; e^{2i\delta_{12}}=det \mathcal{S}_{12}=-\frac{T}{T^{*}}=\frac{\overleftarrow{R}}{\overrightarrow{R}^{*}}\;,
\ee
and they can be related through
\be
e^{2i\delta_{12}}=e^{2i\left(\delta_{1}+\delta_{2}\right)}\frac{1-\left(\overrightarrow{r}_{2}\overleftarrow{r}_{1}\right)^{*}}{\overrightarrow{r}_{2}\overleftarrow{r}_{1}-1}
\label{FP_argument_def}
\ee
which, using the definitions
\be
e^{i\varphi}\equiv\frac{1-\left(\overrightarrow{r}_{2}\overleftarrow{r}_{1}\right)^{*}}{\overrightarrow{r}_{2}\overleftarrow{r}_{1}-1}=\frac{1-\zeta^{*}}{\zeta-1}\; \, ,
\label{FP_argument}
\ee
where $\zeta\equiv\overrightarrow{r}_{2}\overleftarrow{r}_{1}$, can be rewritten under the simple form, 
\be
\delta_{12}-(\delta_{1}+\delta_{2})=\frac{\varphi}{2}.
\label{FP_argument_def2}
\ee
Through $\zeta$, two Fabry-Perot parameters are naturally introduced into the formalism. From  (\ref{FP_argument_def}), we have $\zeta=r_{1}r_{2} {\exp} \left( i \overleftarrow{\theta}_{1}+ i \overrightarrow{\theta}_{2} \right) \equiv\mathcal{{R}}e^{i\theta_{cav}}$, where $\theta_{cav}$ has been defined in (\ref{thetacav}) and $\mathcal{R}(k) \equiv r_{1}r_{2}$ is the mutual reflectance of the substructures, related to the finesse of the resultant  Fabry-Perot cavity by
\be
\mathrm{Finesse}=\frac{\pi\sqrt{\mathcal{{R}}}}{1-\mathcal{{R}}}.
\ee
Thus, we can rewrite the relation between the phase $\varphi$ and the cavity parameters $( \mathcal{\mathcal{{R}}},\,\theta_{cav} )$ as

\be
\begin{array}{c}
\mathcal{\mathcal{{{R}}}}\cos\left(\theta_{cav}+\frac{\varphi}{2}\right)=\cos\left(\frac{\varphi}{2}\right).\end{array}
\label{trigo1}
\ee
Equation (\ref{trigo1}) can be rewritten for nonzero $\mathcal{R}$ as
\be
\frac{\varphi}{2}=\arctan\left(\frac{\mathcal{\mathcal{{{R}}}}\cos\left(\theta_{cav}\right)-1}{\mathcal{\mathcal{{{R}}}}\sin\left(\theta_{cav}\right)}\right). 
\label{trigo2}
\ee
Using relations (\ref{FP_argument}), (\ref{trigo1}) and (\ref{trigo2}), we are now in a position to predict the occurence of gap states. The condition for the appearance of a single gap state $k_{m}$ in a perfect gap ($\mathcal{\mathcal{{{R}}}}=1$) of the spectrum of $[J_{1}J_{2}]$ is 
\be
\delta_{12}(k_{m})\!-\!\left[\delta_{1}(k_{m})+\delta_{2}(k_{m})\right]\!=\!\pi+2\pi m\;, 
\ee
where  $ m \!\in \!\mathbb{Z}$.
Using (\ref{FP_argument}), this condition can equivalently be written as $\varphi\!=\!2\pi m$ and it may be viewed as resulting from the Levinson theorem \cite{Ma2006}. We may rephrase the condition as: a new gap state will arise at  frequencies for which $\varphi$ completes a full winding and the mutual reflectance is perfect. However, this relation is only approximate for the more realistic case $\mathcal{\mathcal{{{R}}}}<1$. From (\ref{trigo1}), one can see that for the perfect mutual reflectance case, $\varphi\left(\theta_{cav},\, \mathcal{\mathcal{{{R}}}}=1\right)=-\theta_{cav}$, which yields the condition for the appearance of a new state, $\theta_{cav}\left(k_{m}\right)=2\pi m$, i.e. nothing but the Fabry-Perot resonance condition (\ref{FP_res2gen}). In Fig. \ref{varphi_tc}, we have studied the effect of varying $\mathcal{R}$ on the relation between $\varphi$, and $\theta_{cav}$ as given by  (\ref{trigo2}). We note that $\varphi$ no longer covers the interval $[0,2\pi]$ and, consequently never takes the (perfect reflection) resonant values as $\theta_{cav}$ does. Instead, it becomes increasingly smeared as $\mathcal{R}$ decreases. 
This shows that in general, the Fabry-Perot parameter $\theta_{cav}$ is indeed more suitable than $\varphi$ to identify the appearance of new gap states in structures with imperfect reflection.

In the next section, we apply the effective Fabry-Perot winding condition (\ref{FP_res2gen}) to the case of topological gap states in quasiperiodic structures. We show that in addition to predicting the resonant frequencies as discussed, the effective Fabry-Perot model allows to understand the spatial symmetry of the resonant modes as being driven by the parity of the Fabry-Perot integer $m$ in (\ref{FP_res2gen}).
In addition and to conclude this section, the effective Fabry-Perot model allows to have a unified description of all gap states producing schemes. Through this "geometrical" understanding, defect and surface states have the same physical origin. 


\section{Topological edge states in quasi-periodic structures}\label{Model_demonstration}
\subsection{The Fibonacci quasicrystal}
The probing of topological edge states is a convenient and largely used method to exhibit  topological properties, either structural or spectral, of a given material \cite{hasan2010,witten2015}.
An important asset of a large class of quasicrystals and quasi-periodic structures is that they have a rich topological content which starts to be systematically unveiled \cite{levy2015,Kraus2012}. Here, we wish to investigate part of these topological features using the properties and the behaviour of conveniently created gap states obtained from  the generalised effective Fabry-Perot condition (\ref{FP_res2gen}). Such edge states, localised at a boundary, have been observed in quasi-periodic structures  \cite{Zijlstra1999,pang2010} and have been analysed in terms of their topological content both in the tight binding approximation \cite{Kraus2012} and by means of a scattering approach \cite{levy2015}. 

\begin{figure}
\includegraphics[viewport=-155bp 1bp 1138bp 321bp, clip,width=1\columnwidth]{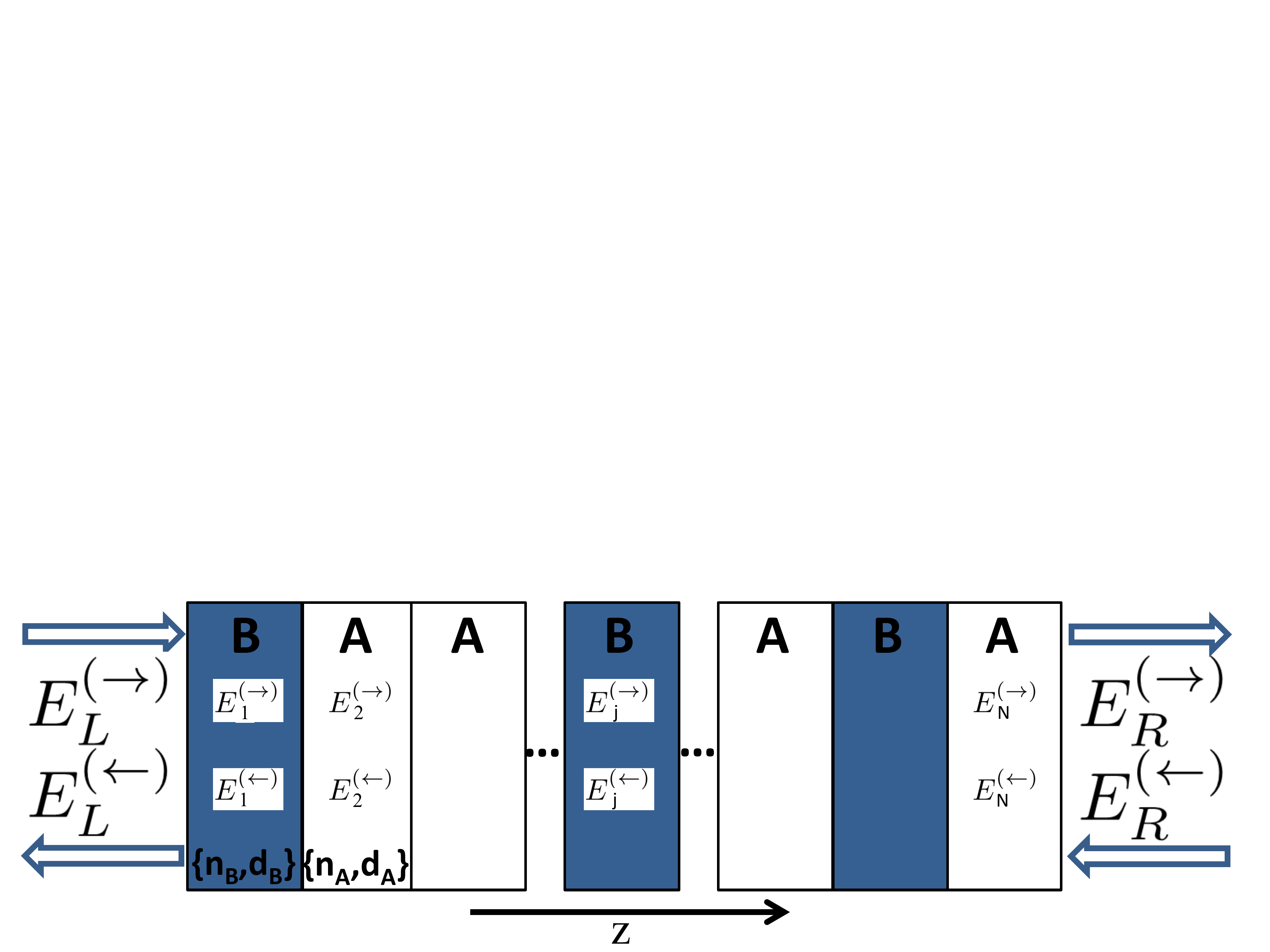}
\caption{(color online) Setup and notations for the dielectric scattering problem. The alphabet $\{A,B\}$ is characterised by a refractive index $\{n_A,n_B\}$ and a thickness $\{d_A,d_B\}$. The notations for the waves are as follows. $\{i_L, o_L\} \equiv \{E_L^{( \rightarrow )} , E_L^{( \rightarrow )}\}$ denote the electric field amplitudes of incoming and outgoing waves at the left boundary, respectively. Similarly, $\{i_R, o_R\} \equiv \{E_R^{( \rightarrow )} , E_R^{( \rightarrow )}\}$ are defined outside the right structure boundary and $\{E_j^{( \rightarrow )} , E_j^{( \rightarrow )}\}$ is defined inside the dielectric structure.\label{notations_dielectric}}
\end{figure}

\begin{figure}
\includegraphics[viewport=-35bp 16bp 698bp 401bp, clip,width=1\columnwidth]{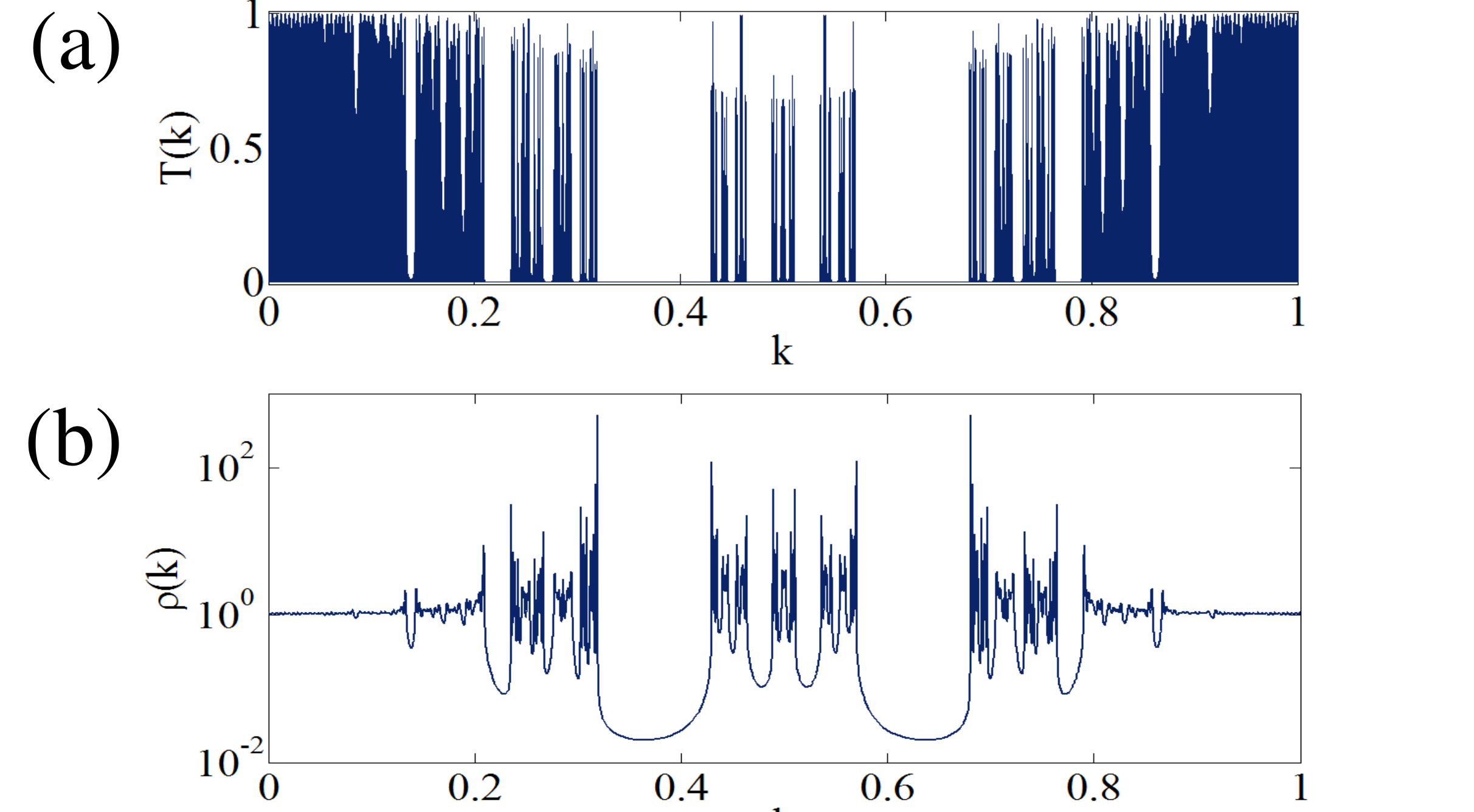}
\caption{(color online) The transmittance ($T\equiv tt^*$) spectrum (a) and the DOS spectrum (b) for the Fibonacci segment $\overrightarrow S_{10}$ as a function of the incoming waves wavenumber $k$.\label{Fibo_spec}}
\end{figure}

A simple version of the Fibonacci quasicrystalline chain is defined from a two letters alphabet $\{A,B\}$. The chain may be constructed by means of several equivalent iterative rules (see \cite{levy2015} and references therein). Here we choose the characteristic function $\chi_m$ which takes the two values $\pm 1$ respectively identified to the letters $ \{ B,A \}$, namely,
\be 
\chi_m = sign \, \left[ \cos \left( 2 \pi \, m \, \tau^{-1} + \phi \right) - \, \cos \left( \pi \, \tau^{-1} \right) \right], 
\label{chi}
\ee
where $m\!\in\!\mathbb{Z}$ and $\tau = (1 + \sqrt{5} ) /2$ is the golden mean. The angular degree of freedom $ \phi$ can be safely ignored for the infinite chain $S_\infty$ but not for any of its finite segments generally defined by $\overrightarrow{S}_j (\phi)\!\!\equiv\!\! \left[ \chi_1 \, \!\chi_2 \,\! \cdots \! \chi_{F_j} \right]$ (with $F_1=1,\;F_2=2,\;F_{j>2}=F_{j-2}+F_{j-1}$). The alphabet $\{A,B\}$ generally represents a piecewise modulation of a physical parameter (e.g. density, potential, dielectric constant etc.). Here we discuss the case of a dielectric modulation, such that the letters $\{A,B\}$ represent different values of refractive index $\{n_A,n_B\}$, and different layer thickness $\{d_A,d_B\}$, as depicted in Fig. \ref{notations_dielectric}. The DOS and the transmission spectrum of a Fibonacci dielectric structure, calculable using the scattering approach, has a fractal structure \cite{akkermans2013evgeni, akkermans2013,Tanese2014} made out of a rich variety of bands and gaps (see Fig. \ref{Fibo_spec}). 


\subsection{Gap states in a Fibonacci quasicrystal}

\begin{figure}
\includegraphics[viewport=-155bp 16bp 988bp 531bp, clip,width=1\columnwidth]{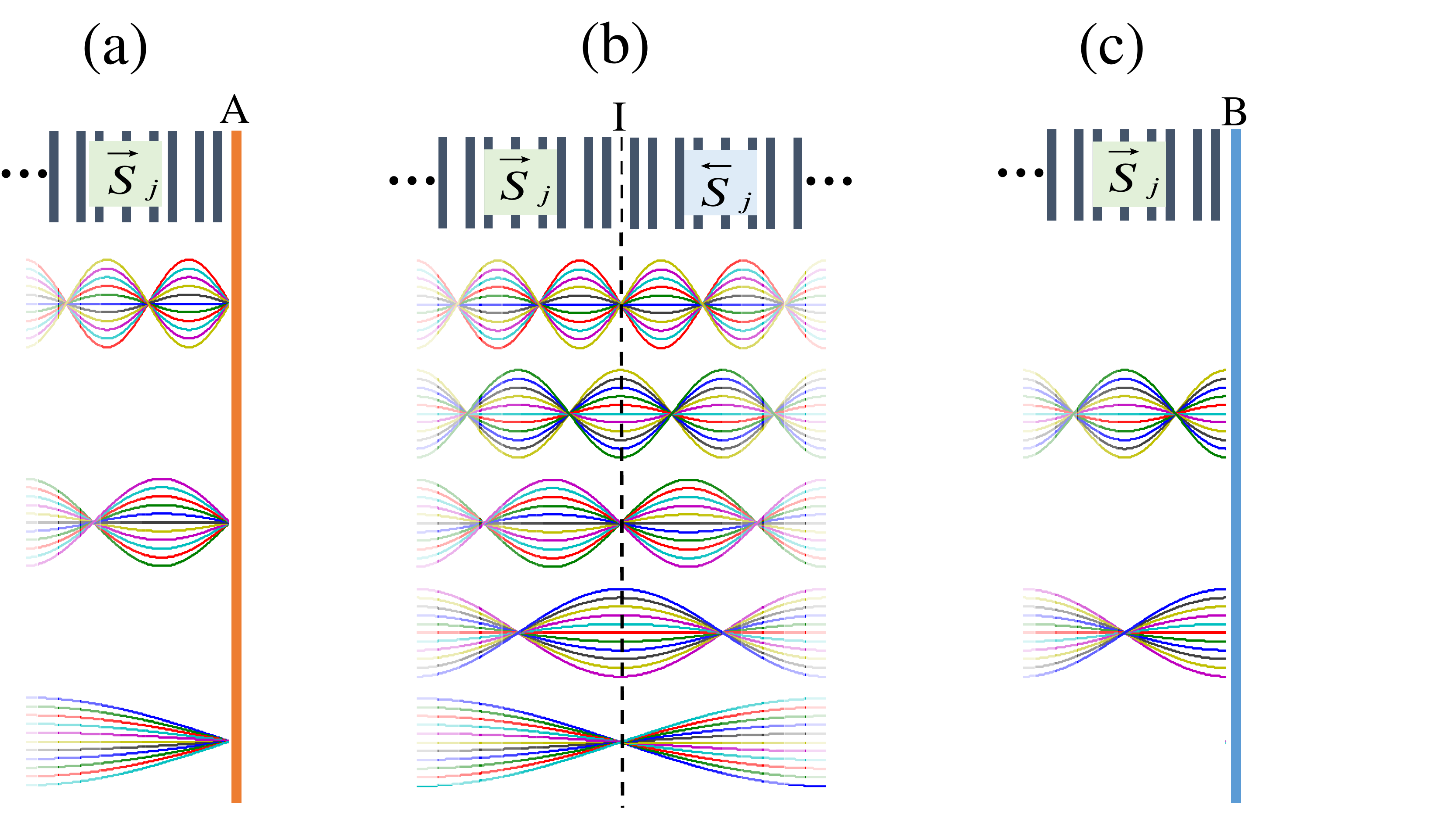}
\caption{(color online) (a) and (c): The Fibonacci chain segment $\overrightarrow S_j$ (indicated by black and white bars) bounded from the right by a metallic mirror (a) indicated by an orange line and marked with ``A'', or by a continuum with refractive index smaller than that of the chain (c) indicated by a blue line and marked with ``B'', hosting gap states localised at the edge as requested by the boundary conditions. (b) The artificial palindrome based on the same structure, hosting gap states localised in the interface indicated by the dotted line and marked with ``I''. The artificial palindrome scheme serves as a generalised mirror hosting gap states of both spatial symmetries.\label{Art_pal}}
\end{figure}

One scheme recently proposed to produce gap states relevant for the topological analysis of quasicrystals considers a symmetrized hetero-structure built out of a chain and its mirror image (see Fig. \ref{Art_pal}.b). To that purpose, consider a given chain bounded from one side by a perfect mirror (either metallic as in Fig. \ref{Art_pal}.a or index mismatch based as in Fig. \ref{Art_pal}.c). 
Waves traveling towards the mirror plane are reflected back into the chain, experiencing the dielectric modulation in a reverse order. An equivalent scattering version of this setup consists in removing the mirror and instead unfolding the chain with respect to the mirror plane. In that way, we create an artificial symmetric chain (an artificial palindrome). From the previous description of a Fibonacci chain and using the notations of section \ref{FP}, we consider the artificial palindrome denoted by $[J_1J_2]=[\protect\overrightarrow{S}_j\protect\overleftarrow{S}_j]$, where $\overleftarrow{S}_j (\phi)\!\!\equiv\!\! \left[ \chi_{F_j} \, \!\chi_{F_j-1} \,\! \cdots \! \chi_1 \right]$. Note that symmetry-wise, the possible gap states spatially localised around the interface of the artificial palindrome are equivalent to the union of the possible edge states in the metallic mirror case (with a node in the mirror plane/interface), and the possible edge states in the mismatch mirror case (with an anti-node in the mirror plane/interface). Therefore, the artificial palindrome scheme is a generalised mirror/edge, hosting twice the number of gap states as compared to any type of mirror. Noting that in the artificial palindrome case
\be
\delta_1=\delta_2\equiv\delta\,;\,\overleftarrow\theta_1=\overrightarrow\theta_2\equiv\overleftarrow\theta\,;\,\overrightarrow\theta_1=\overleftarrow\theta_2\equiv\overrightarrow\theta
\ee
and using (\ref{scatt_total_ps}) and (\ref{scatt_chiral_phase}), the relevant scattering phase (\ref{thetacav}) for the artificial palindrome rewrites as, 
\be
\theta_{cav}\left(k,L,\overleftarrow{\theta} ,\overrightarrow{\theta} \right) \!\!=\frac{4\pi L}{\lambda\left(k\right)}+\!2\overleftarrow{\theta}=\frac{4\pi L}{\lambda\left(k\right)}+\!2\delta+\alpha,
\label{thetacav_ap}
 \ee
emphasizing that for the artificial palindrome, the cavity phase is a combination of the independent scattering phases given in (\ref{scatt_phases}). Here we show that the resultant resonant gap states of the Fibonacci quasi-periodic chain are of a geometric nature \cite{levy2015}.

\begin{figure}
\includegraphics[viewport=-55bp 16bp 808bp 301bp, clip,width=1\columnwidth]{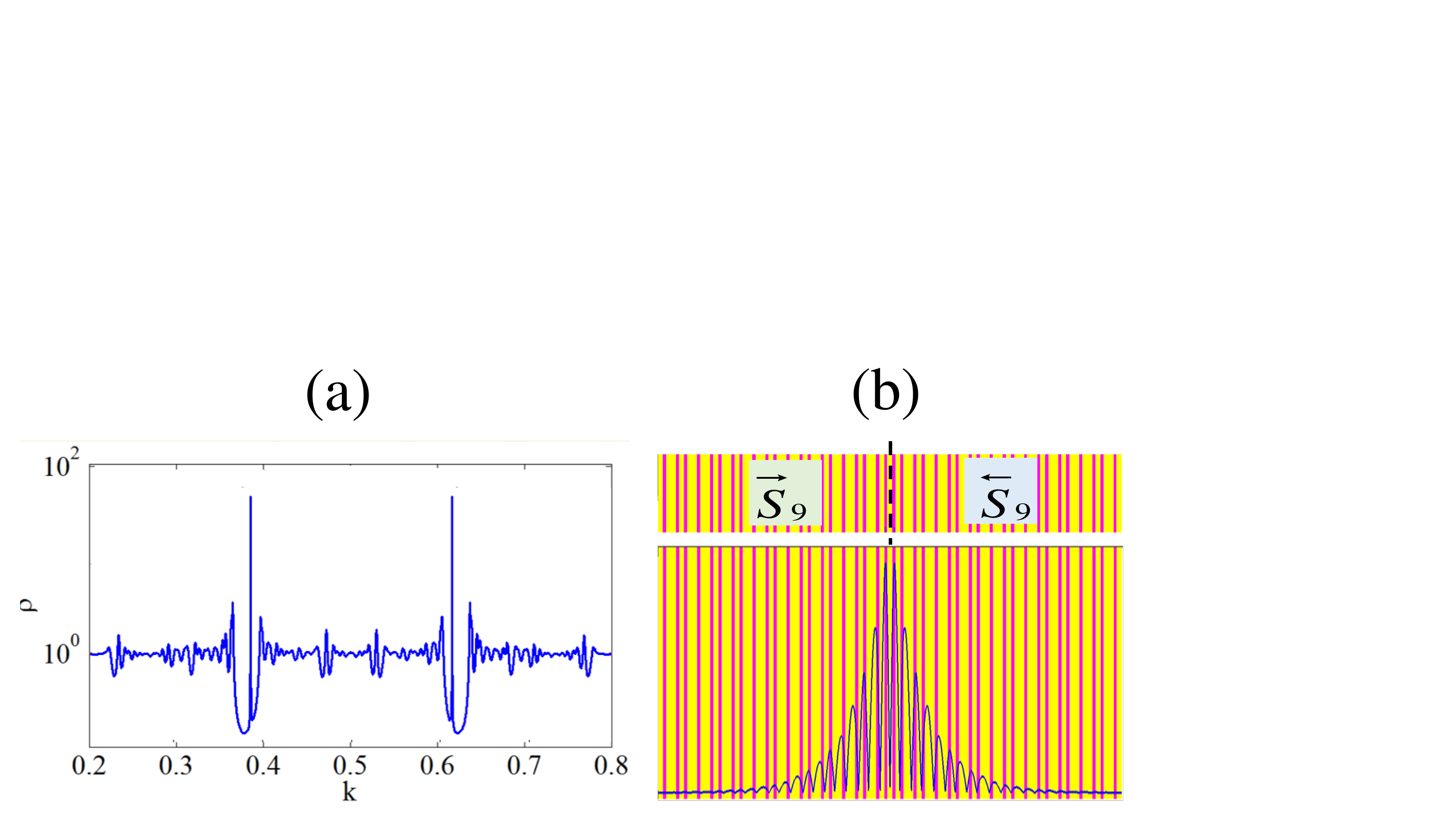}
\caption{(color online) The appearance of gap states in the artificial palindrome scheme for the Fibonacci segment $\overrightarrow S_9$. (a) The DOS spectrum with new states appearing at the gaps. (b) The spatial arrangement of the heterostructure indicated by yellow and magenta bars, with a representation of the electric field intensity for the gap state at $k=0.39$ showing a node at the interface.\label{Fibo_gap_states}}
\end{figure}

\begin{figure}
\includegraphics[viewport=-55bp 2bp 808bp 301bp, clip,width=1\columnwidth]{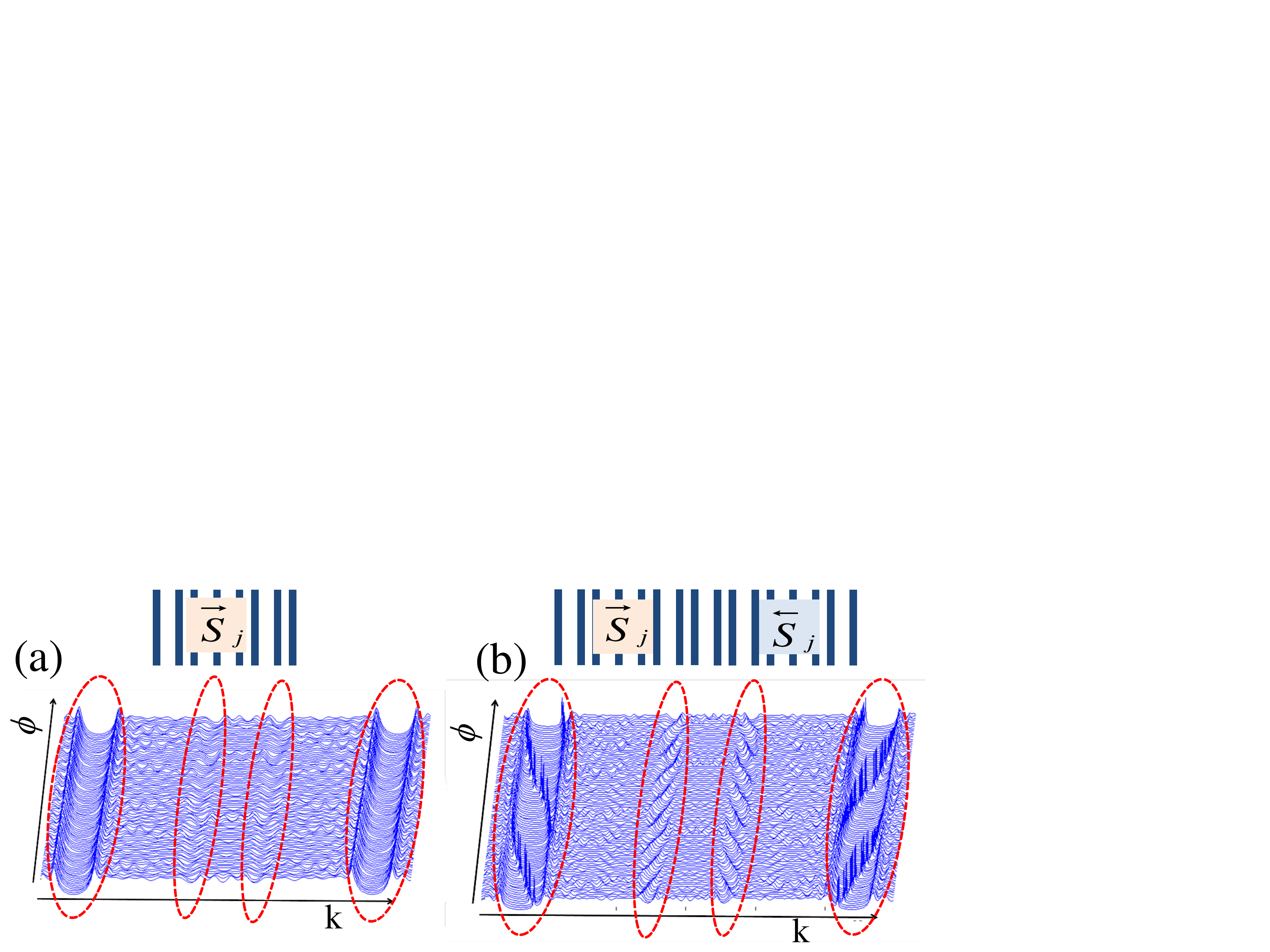}
\caption{(color online) Topological gap states in the spectrum of the artificial palindrome $[\protect\overrightarrow{S}_{10}\protect\overleftarrow{S}_{10}]$. (a) A semi 3D plot of the density of states as a function of $\phi$  and $k$ for the chain $\overrightarrow S_{10}$. The $\phi$-independent gaps are marked with dashed red ellipses. (b) The same as (a), for the artificial palindrome $[\protect\overrightarrow{S}_{10}\protect\overleftarrow{S}_{10}]$. Gap states and their traverse as a function of $\phi$ is observable.\label{Fibo_gap_states_phi}}
\end{figure}

The DOS and transmission spectrum of $[\protect\overrightarrow{S}_j\protect\overleftarrow{S}_j]$ have their bands and gaps distributed as for the single (i.e. mirrorless) structure. Gap states are spatially localized around the heterostructure interface \cite{levy2015} (see Figs. \ref{Fibo_gap_states}, and \ref{Fibo_gap_states_phi}). 
Figure \ref{Fibo_gap_states_phi} shows that scanning the angular degree of freedom $ \phi$ in (\ref{chi}) does not change the spectral locations of neither the bands nor the gaps, but significantly affects  their spectral location. Such a behaviour has been shown to be directly related to the topological (Chern) invariants ascribed to each gap \cite{levy2015, Dareau2016, baboux2016}.

\begin{figure}
\includegraphics[viewport=-55bp 6bp 1008bp 301bp, clip,width=1\columnwidth]{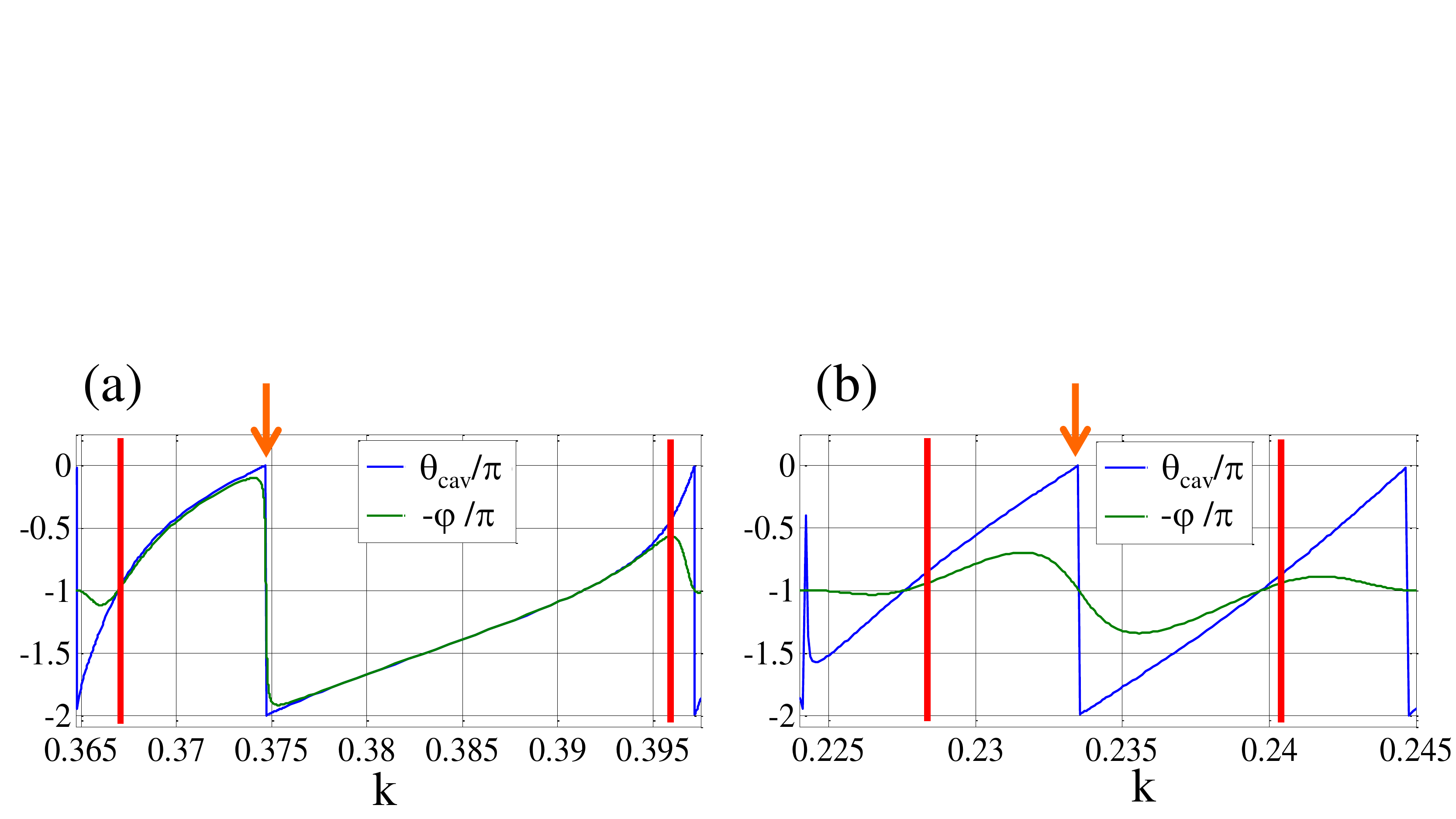}
\caption{(color online) The behavior of the phases $\varphi(\theta_{cav},\mathcal{R})$ and $\theta_{cav}$ as a function of the incoming waves wavenumber $k$ for the artificial palindrome based on the Fibonacci segment $\overrightarrow S_{10}$, at two different spectral gaps (gap edges are indicated by red bars, and the gap state frequency is indicated by orange arrows). (a) Gap around $k= 0.385$, with $\mathcal{R}=r^2=0.98$. (b) Gap around $k= 0.235$, with $\mathcal{R}=0.25$.\label{Fibo_vp_tc}}
\end{figure}

\begin{figure}
\includegraphics[viewport=-155bp 10bp 728bp 391bp, clip,width=1\columnwidth]{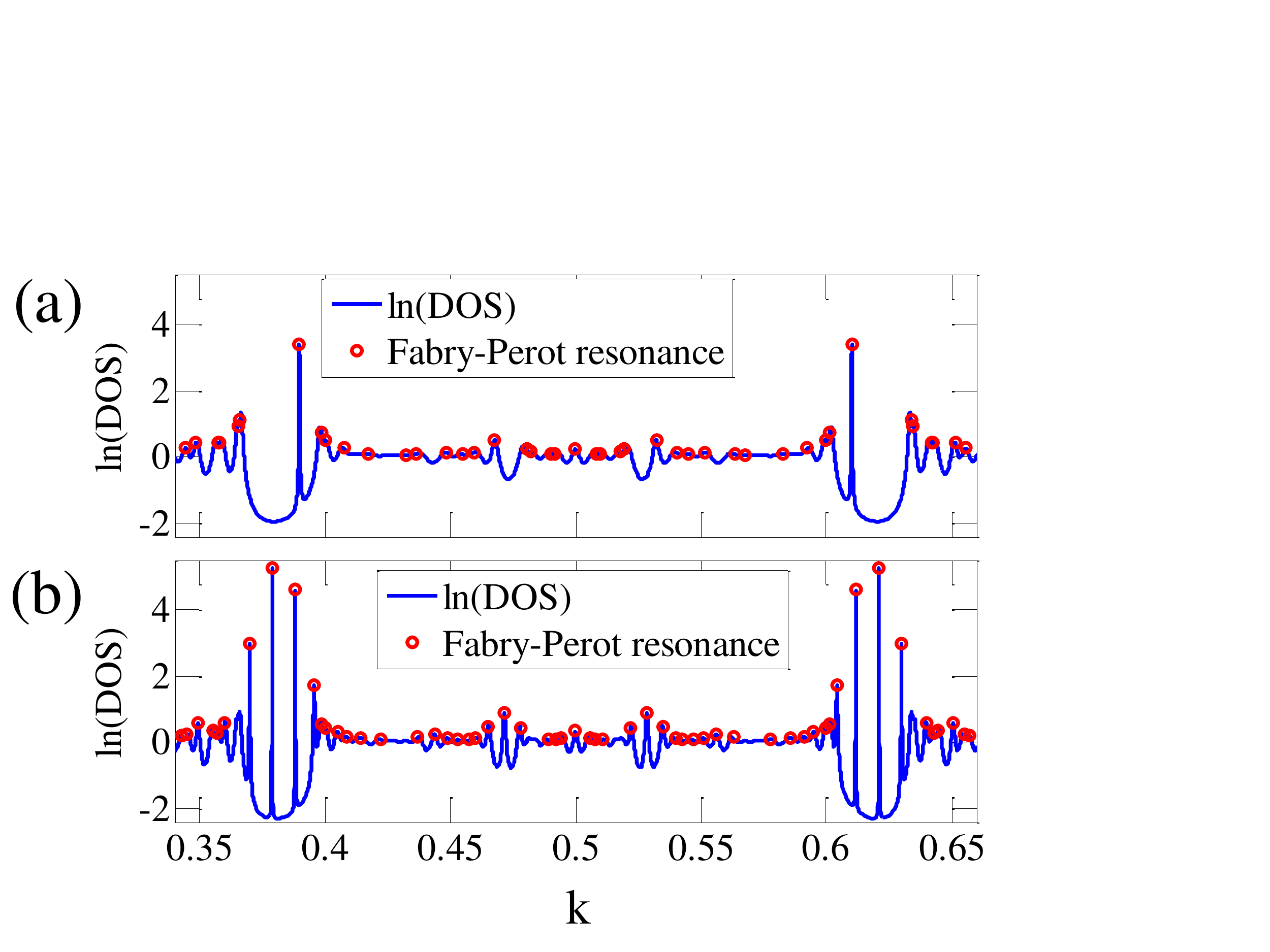}
\caption{(color online) (a) DOS spectrum of the artificial palindrome based on the Fibonacci segment $\overrightarrow S_{10}$. Red circles mark the Fabry-Perot resonant wavevectors $k_m$ according to  (\ref{FP_res2gen}). Resonances within the spectral gaps are the subject of this note, while the resonances lying within the bands are related to the so called perfect transmission resonances. (b) The geometrical cavity setup: the substructures $\overrightarrow S_{10}$, and $\overleftarrow S_{10}$ are separated by a region with a uniform refractive index (with a geometrical cavity length approximately as long as $\overrightarrow S_{10}$). The DOS spectrum (blue lines) and the resonant condition prediction (red circles) are indicated.\label{gapmodes_prediction}}
\end{figure}

\begin{figure}
\includegraphics[viewport=25bp 100bp 938bp 481bp, clip,width=1\columnwidth]{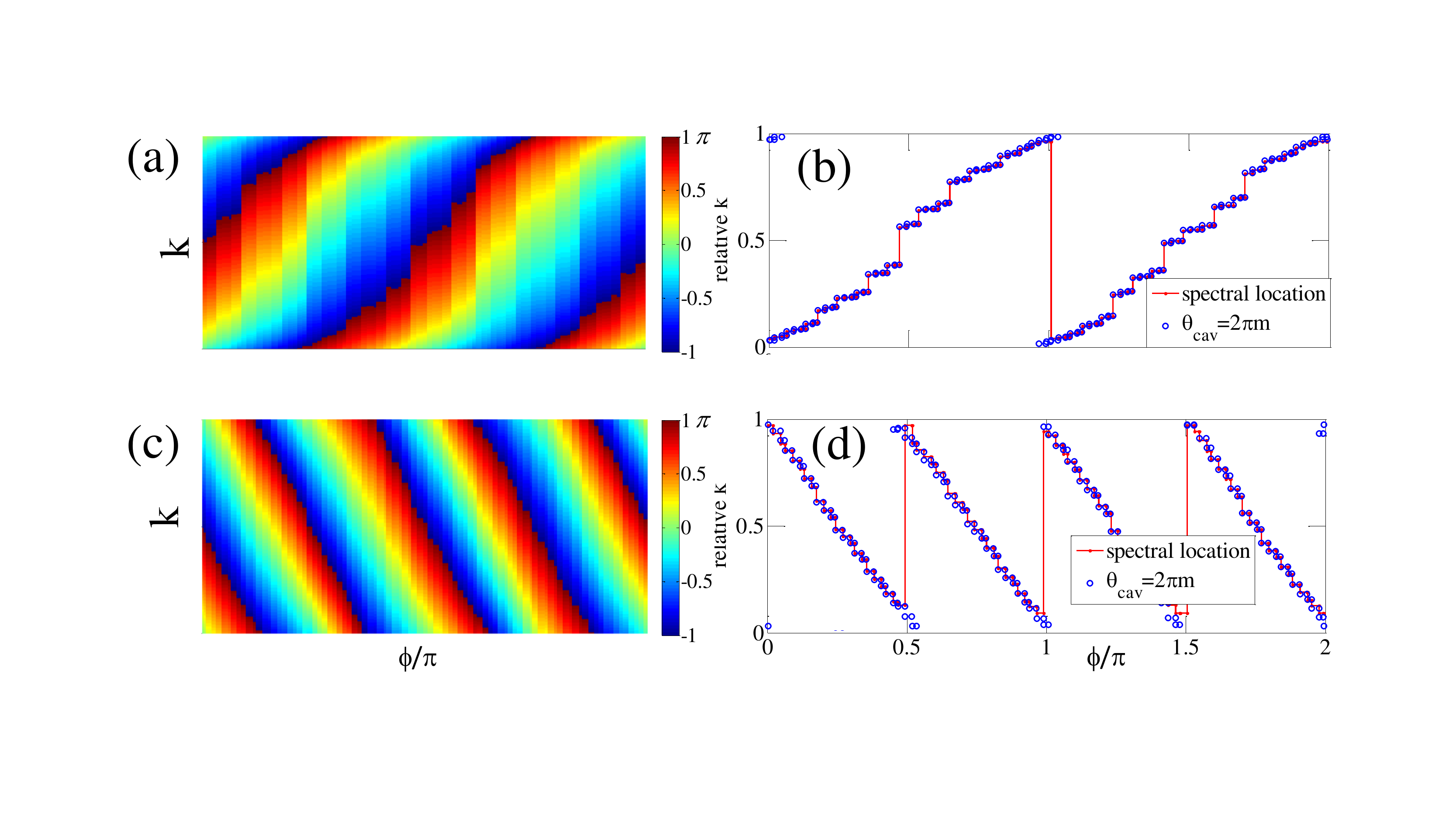}
\caption{(color online) Topological gap states in the spectrum of the artificial palindrome based on the Fibonacci segment $\overrightarrow S_{10}$. (a) Colormap of $\theta_{cav}(\phi,k)$ within the gap around $k= 0.385$. Light green areas correspond to the Fabry-Perot resonance condition (\ref{FP_res2}). (b) Crossover of the gap states as a function of $\phi$. The relative spectral location of gap states (red) is compared to the Fabry-Perot resonance condition (blue).(c)-(d) The same as (a)-(b) for the gap around $k= 0.235$.\label{gapmodes_tc}}
\end{figure}

Now, we show how to obtain and characterise these states using the effective Fabry-Perot model (\ref{FP_res2gen}) (see also Fig. \ref{Fibo_vp_tc}). For a given value of $\phi$, one can calculate the phases $\varphi(k)$ and $\theta_{cav}$  (discussed before and depicted in Fig. \ref{varphi_tc}) for each relevant spectral gap. The behavior of $\theta_{cav}\!=\!2\overleftarrow{\theta}$, and $\varphi\!=\!2\delta_{12}\!-\!4\delta$ is examined for two selected spectral gaps, characterized by different values of $r^{2}$. Indeed, the winding range of $\theta_{cav}$ is unaffected by values of the reflectance $r^{2}$ different from unity, while $\varphi$ ceases to cover the interval $[0,2\pi]$ (even for $r^{2}=0.98$, Fig.\ref{Fibo_vp_tc}.a). Thus, for
any value of $r^{2}$, the use of the condition (\ref{FP_res2})  to calculate the gap state frequencies for $\mathcal{P}$ is justified. Specifically, if this condition is met within the spectral gap (orange arrows in Fig.\ref{Fibo_vp_tc}), then a gap state is expected to appear at this frequency. The prediction power of this process is demonstrated in Fig. \ref{gapmodes_prediction}a. Later on, we shall see that the appearance of gap states in single segments (no symmetrization) with a reflective boundary condition can also be predicted using this exact process with an additional selection rule required to meet the boundary condition. 
To illustrate the geometrical origin of these topological gap states, we introduce a geometrical cavity in addition to the artificial palindrome virtual cavity (i.e. the sub-structures are now separated by a uniform region of length $L$). This leads to a decreasing spectral distance between successive Fabry-Perot resonances, and thus to the existence of more than one resonance within a single gap as displayed in Fig. \ref{gapmodes_prediction}b.
In the topological characterisation of these gaps using the winding of the corresponding phase $\theta_{cav}$  \cite{levy2015}, this process is repeated for all values of $\phi$ to produce a gap state trajectory as a function of $\phi$. This result for two selected spectral gaps is displayed in Fig. \ref{gapmodes_tc}. 

\begin{figure}
\includegraphics[viewport=-55bp 16bp 1038bp 491bp, clip,width=1\columnwidth]{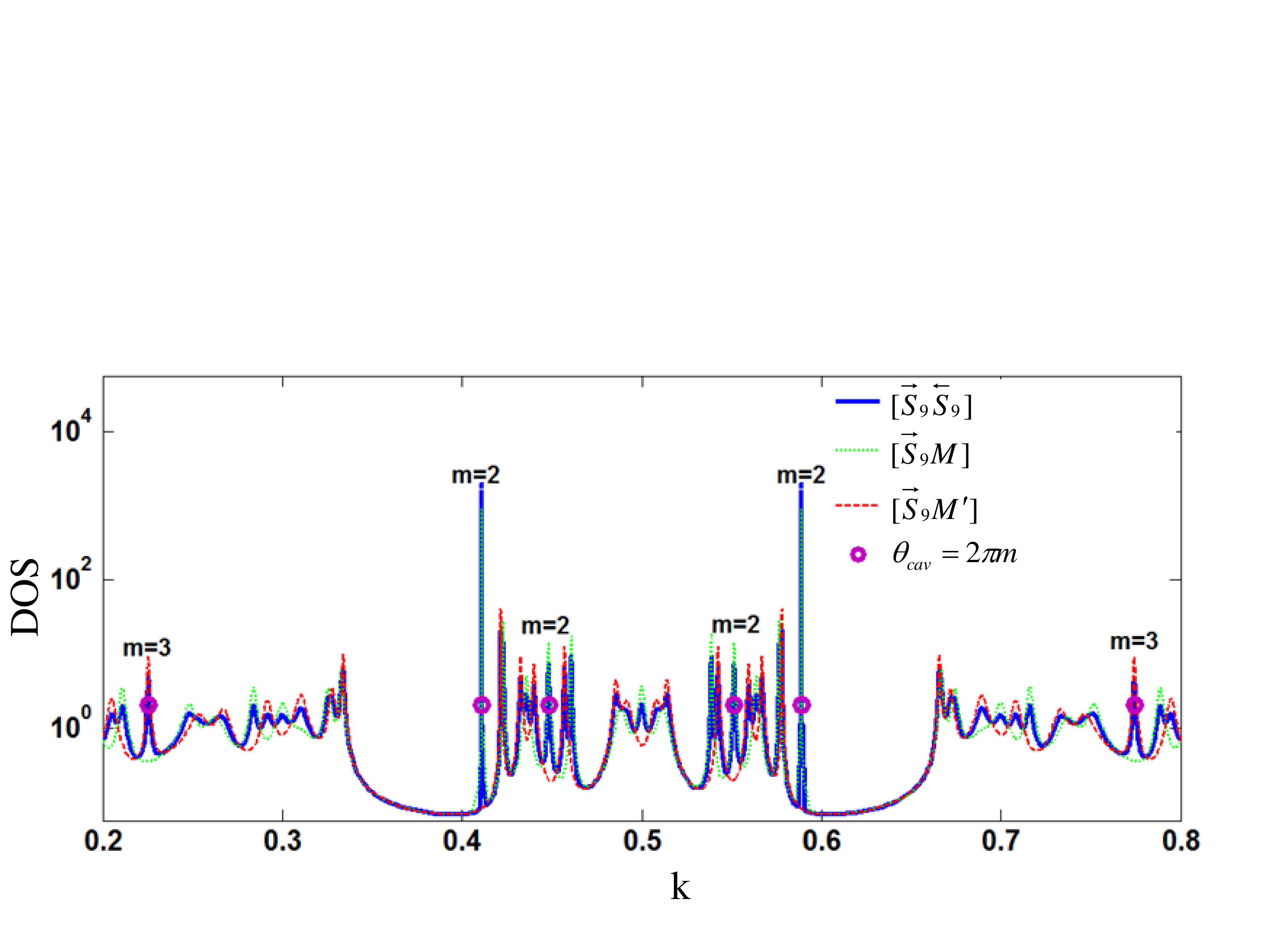}
\caption{(color online) The Fibonacci chain with a different choice of boundary conditions. The DOS spectrum for $\overrightarrow S_9$ based artificial palindrome  is represented (in blue), and for a metallic (dashed red) and a refractive index mismatch (dashed green) reflective boundary conditions on the right edge. The effective Fabry-Perot model intragap solutions $k_m$ are depicted by purple circles.\label{Fibo_gapmodes_spec}}
\end{figure}

\begin{figure}
\includegraphics[viewport=-155bp 16bp 988bp 541bp, clip,width=1\columnwidth]{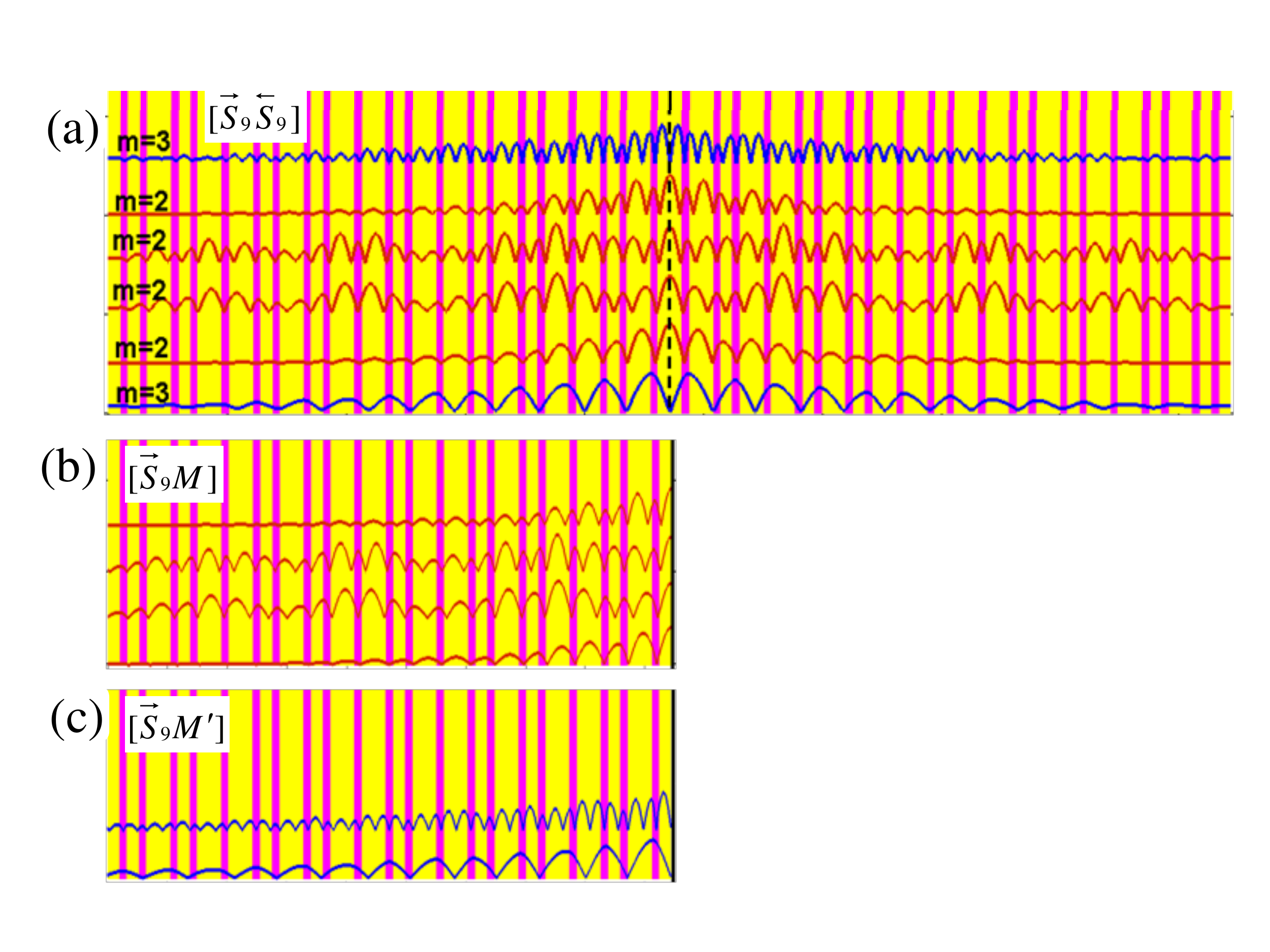}
\caption{(color online) The electric field intensity profile for selected gap states of the structures of Fig. \ref{Fibo_gapmodes_spec}: (a) The $\overrightarrow S_9$ based artificial palindrome. (b) The
$\overrightarrow S_9$ chain with a refractive index mismatch boundary condition on the right edge. (c) The same as (b) for a metallic reflective boundary.\label{Fibo_gapmodes_spat}}
\end{figure}

\begin{figure}
\includegraphics[viewport=-55bp 16bp 1038bp 741bp, clip,width=1\columnwidth]{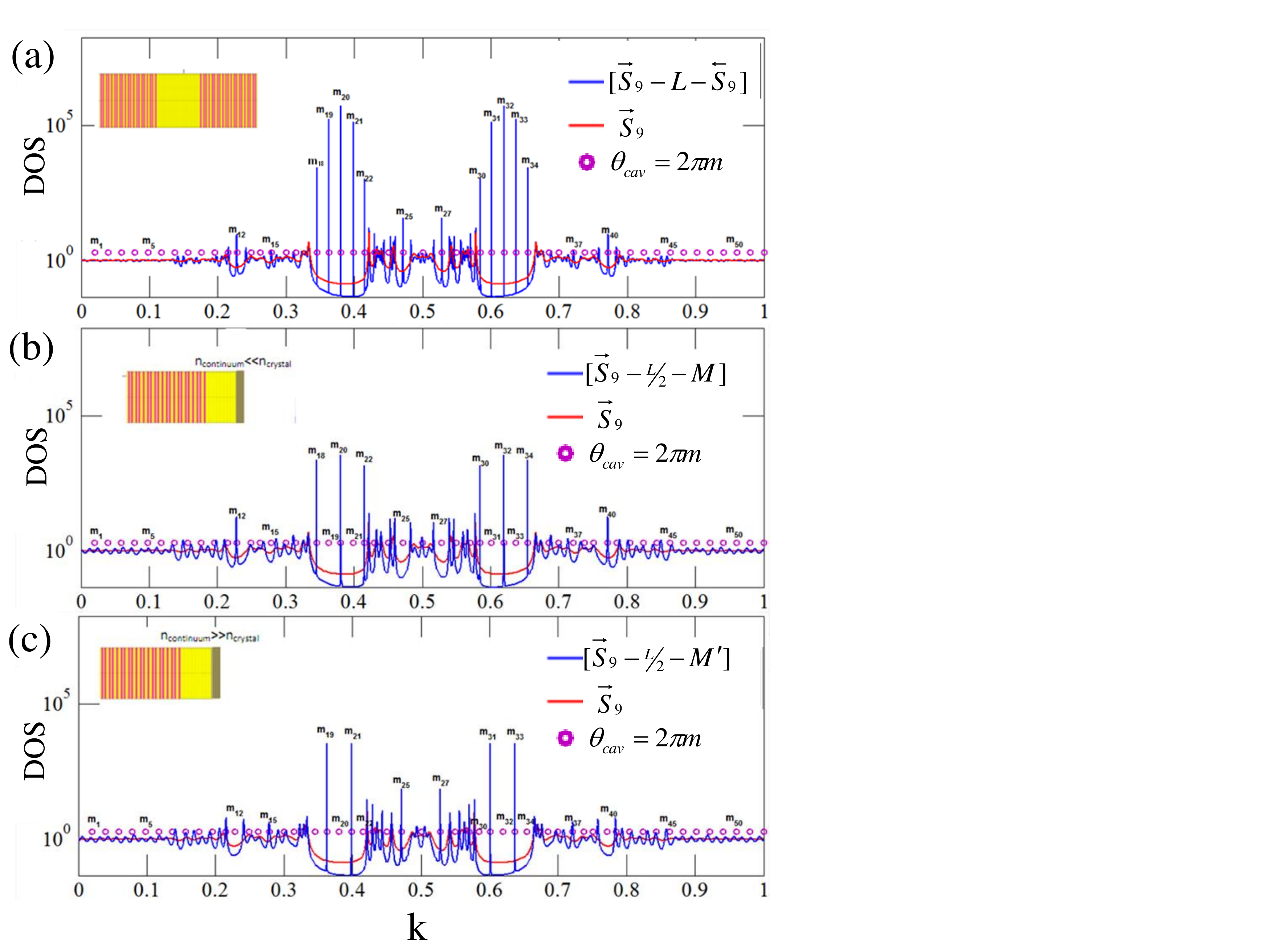}
\caption{(color online) The same as Fig. \ref{Fibo_gapmodes_spec} with an additional region with a uniform refractive index. (a) The artificial palindrome case with a geometrical cavity $L$ (in blue), compared to the unperturbed structure (in red). The effective Fabry-Perot model intra-gap solutions $k_m$ are depicted by purple circles. (b) The same as (a) for the refractive index mismatch boundary condition with an $L/2$ stand-off. (c) The same as (b) for a metallic reflective boundary.\label{Fibo_gapmodes_spec2}}
\end{figure}

\begin{figure}
\includegraphics[viewport=-55bp 16bp 1038bp 741bp, clip,width=1\columnwidth]{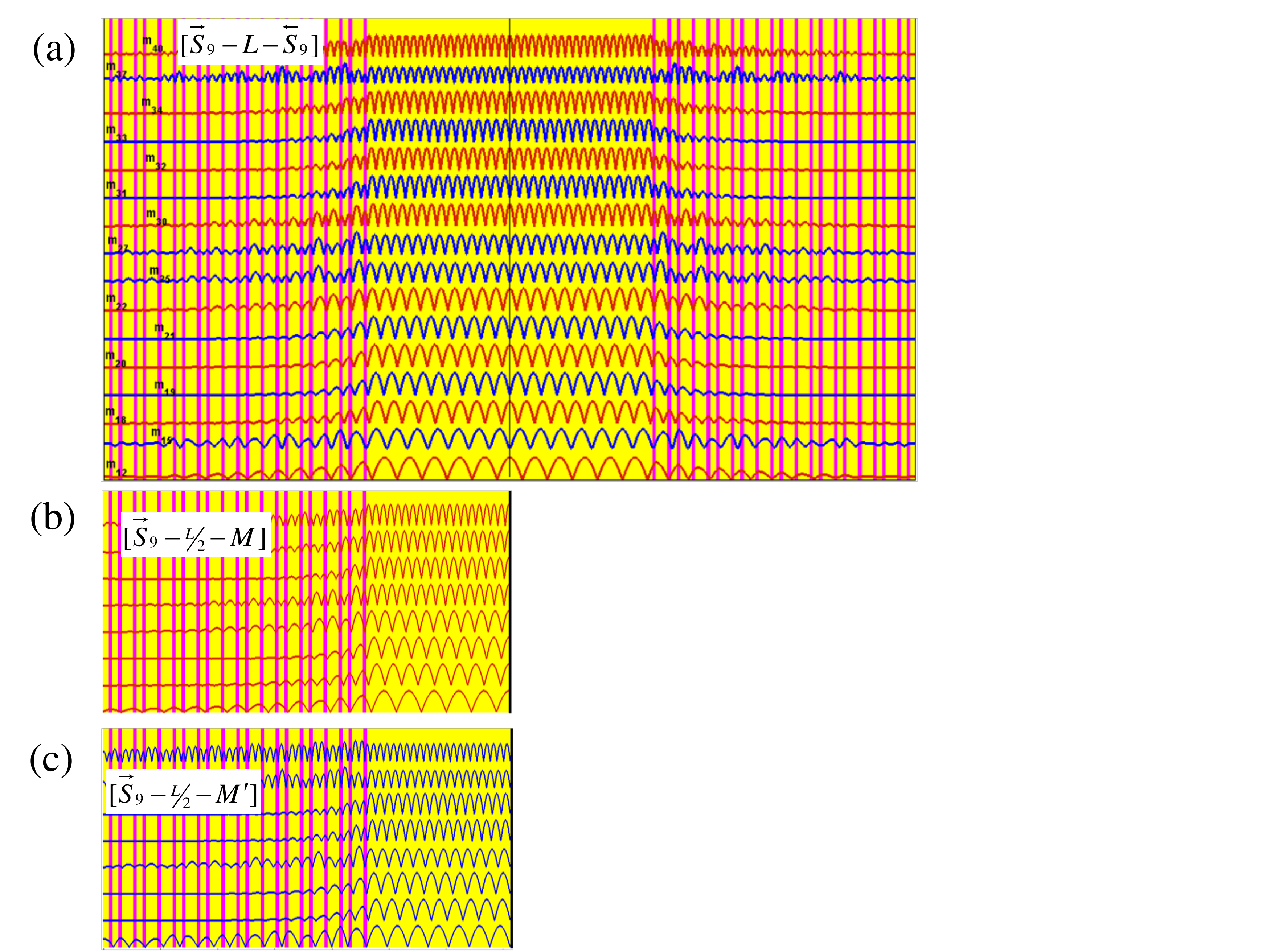}
\caption{(color online) The electric field intensity profile and spatial symmetry for selected gap states of the structures of Fig. \ref{Fibo_gapmodes_spec2}. (a) The $\overrightarrow S_9$ based artificial palindrome with a geometrical cavity $L$. (b) The $\overrightarrow S_9$ chain  with a refractive index mismatch boundary condition on the right edge with an $L/2$ standoff to the structure. (c) The same as (b) for a metallic reflective boundary.\label{Fibo_gapmodes_spat2}}
\end{figure}

As described earlier, the use of the effective Fabry-Perot model for heterostructures with many spectral gaps,  differs significantly from the usual Fabry-Perot resonance ``comb'' picture. However, besides predicting the resonant frequencies, more information may be extracted about the spatial properties of the gap states through the parity of the Fabry-Perot integer, $m$ in (\ref{FP_res2gen}). In the usual Fabry-Perot picture for phase conserving mirrors (a condition such that the refractive index of the outside world is significantly smaller than the refractive indices $n_A$ and $n_B$), odd $m$ leads to  anti-symmetric states and even $m$ leads to symmetric states (with respect to the mid-cavity coordinate). This result is also true for our generalised virtual Fabry-Perot cavities. Figures \ref{Fibo_gapmodes_spec} and \ref{Fibo_gapmodes_spat} shows that the parity of $m$ accurately predicts the existence of a node or an anti-node in the mid-cavity coordinate for the virtual cavity setup. Figures \ref{Fibo_gapmodes_spec2} and \ref{Fibo_gapmodes_spat2} present the same behavior for the case of an additional geometric cavity. In addition, figures \ref{Fibo_gapmodes_spec}-\ref{Fibo_gapmodes_spat2}  illustrate the geometric origin of boundary states either at the interface of the artificial palindrome and of edge states in the case of reflective boundary conditions. The artificial palindrome indeed plays the role of a generalized edge, hosting gap states of both spatial symmetries (with respect to the mid-cavity coordinate). This additional characterisation of gap states has been proposed to probe topological properties of spectral gaps in quasiperiodic chains \cite{baboux2016}.
\begin{figure}
\includegraphics[viewport=-55bp 56bp 1008bp 491bp, clip,width=1\columnwidth]{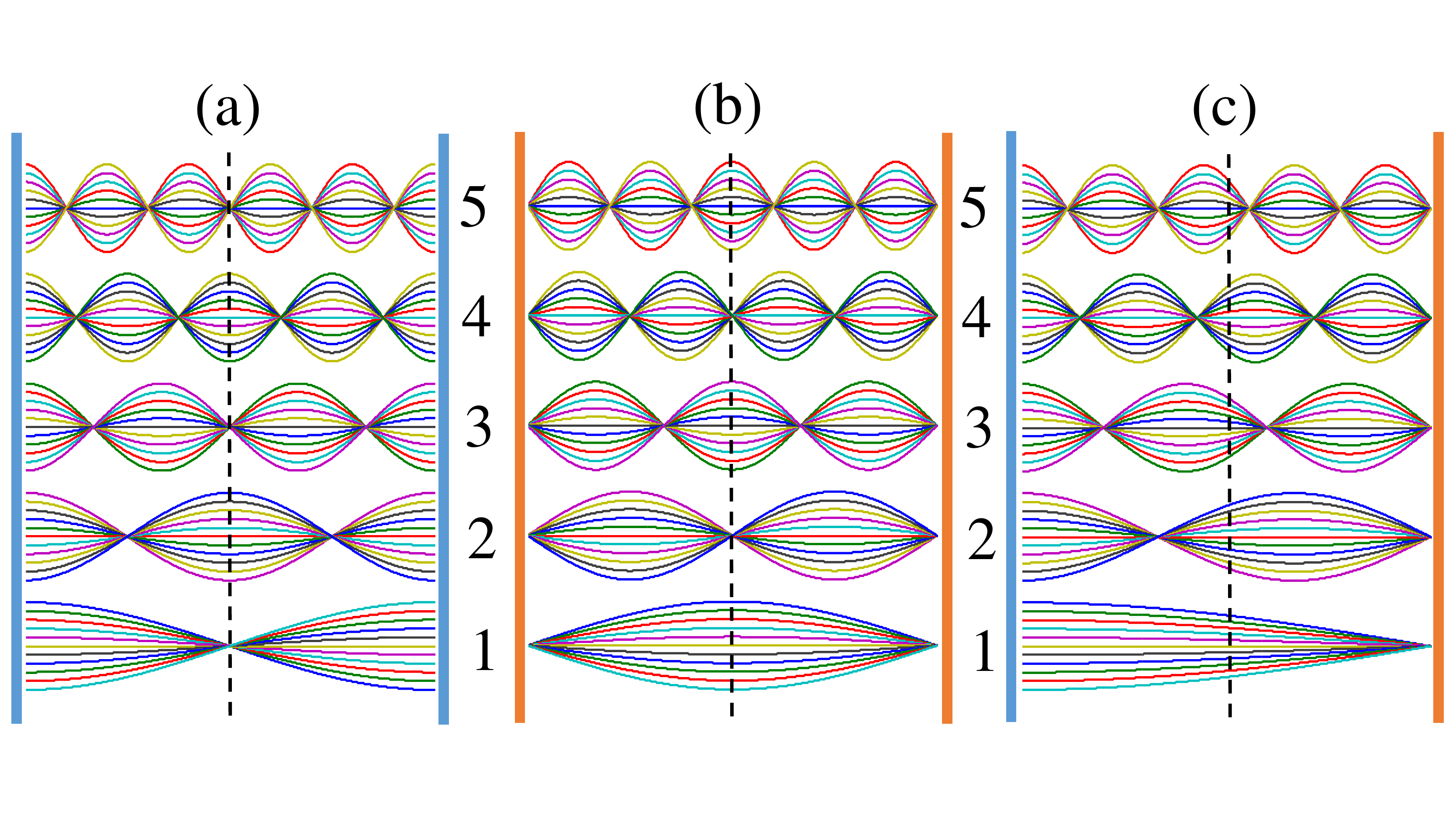}
\caption{(color online) The first 5 Fabry-Perot cavity resonant states (and symmetry) for various
mirror schemes. (a) The case with two index mismatch phase conserving mirrors indicated by blue bars and marked ``B''. (b) The case with two metallic phase flipping mirrors indicated by orange bars and marked ``A''. (c) The hybrid case with a metallic mirror on one side, and an index mismatch mirror on the other.\label{FPsym1}}
\end{figure}

\begin{figure}
\includegraphics[viewport=-55bp 66bp 758bp 341bp, clip,width=1\columnwidth]{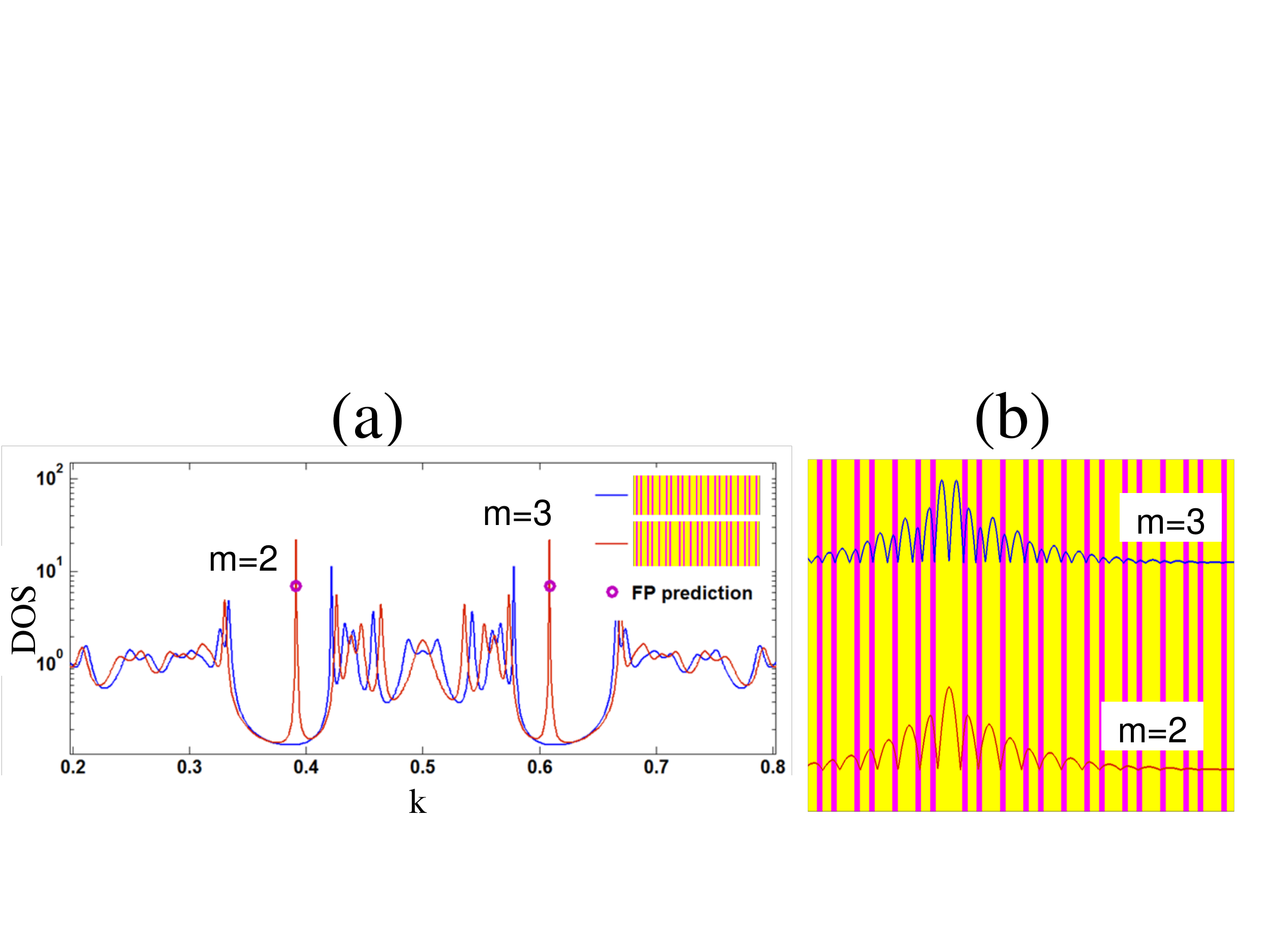}
\caption{(color online) The Fibonacci chain $S_9$ with a non central single letter defect. (a)
DOS spectrum for the unperturbed structure (in blue) compared to that of the same
structure with a single non central defect (in red). The two intra-gap solutions  $k_m$ of the effective Fabry-Perot model with $m = 2, 3$ are in purple circles. (b) The electric field intensity of the defect gap states $m = 2$ (in red) and $m = 3$ (in blue) on top of the structural detail (indicated by yellow and magenta bars).\label{Fibo_defect}}
\end{figure}

To emphasize the generality of this generalised Fabry-Perot approach, and to illustrate the common origin of all gap states, we consider a single Fibonacci segment with a single structural defect (in this case an additional ``$A$'' layer inserted within the structure). The gap states which arise are termed defect states, and are depicted in Fig. \ref{Fibo_defect}. However, if we slice this structure in the middle of the defect and consider the virtual Fabry-Perot cavity defined by the resultant sub-structures, we find that it perfectly predicts the existence and frequency of the defect states, as well as their spatial symmetry (through the parity of $m$).


\subsection{A topological and  wavelength dependent cavity length}\label{WDCL}

\begin{figure}
\includegraphics[viewport=-125bp 16bp 1038bp 541bp, clip,width=1\columnwidth]{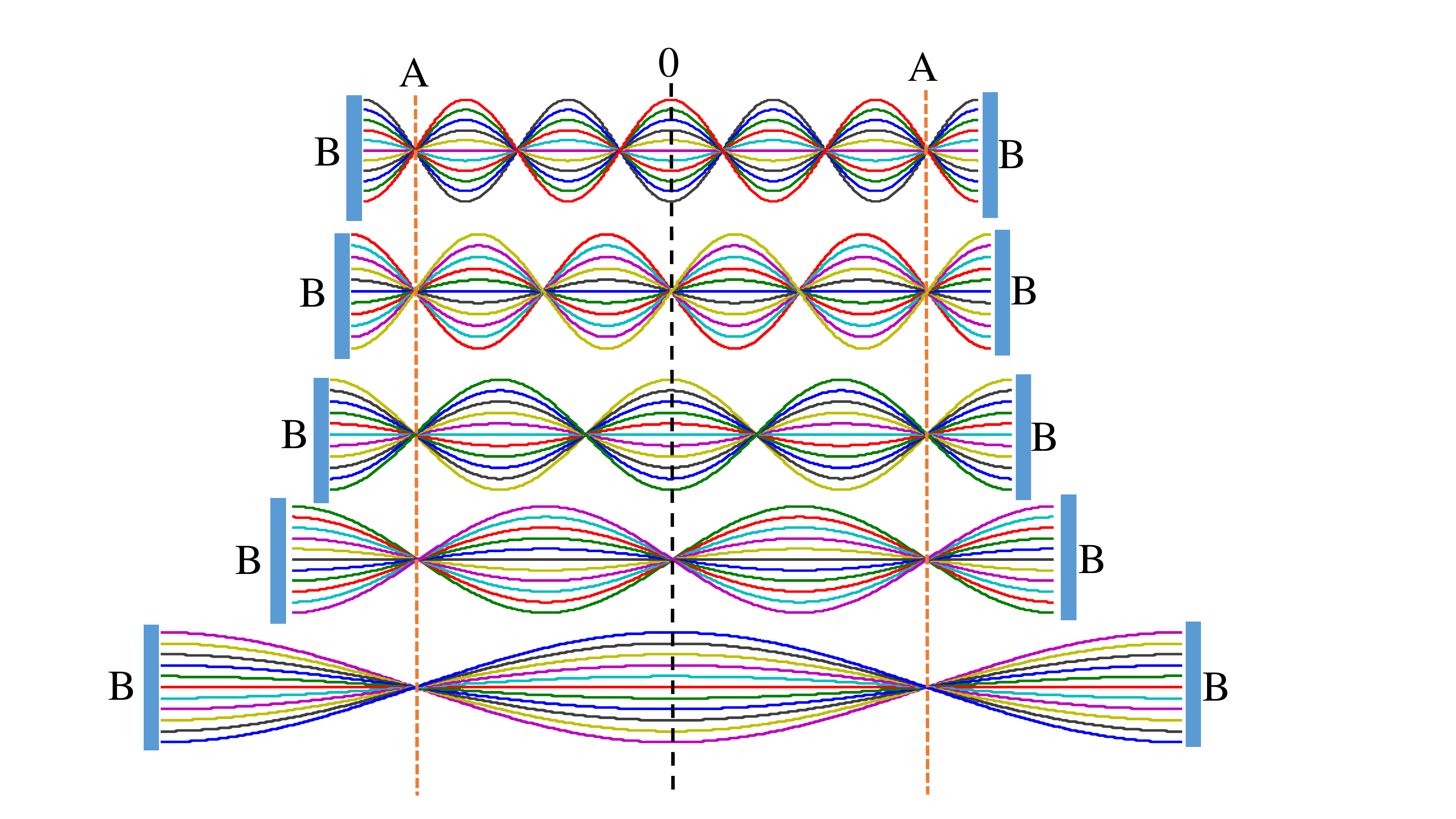}
\caption{(color online) A sketch of case (b) in Fig. \ref{FPsym1} clarified by replacing each metallic mirror (dashed orange lines marked ``A'') by a phase conserving mirror (blue bars marked ``B''), retracted by a quarter wavelength, i.e. a frequency dependent cavity length.\label{FPsym2}}
\end{figure}

\begin{figure}
\includegraphics[viewport=-55bp 16bp 1078bp 531bp, clip,width=1\columnwidth]{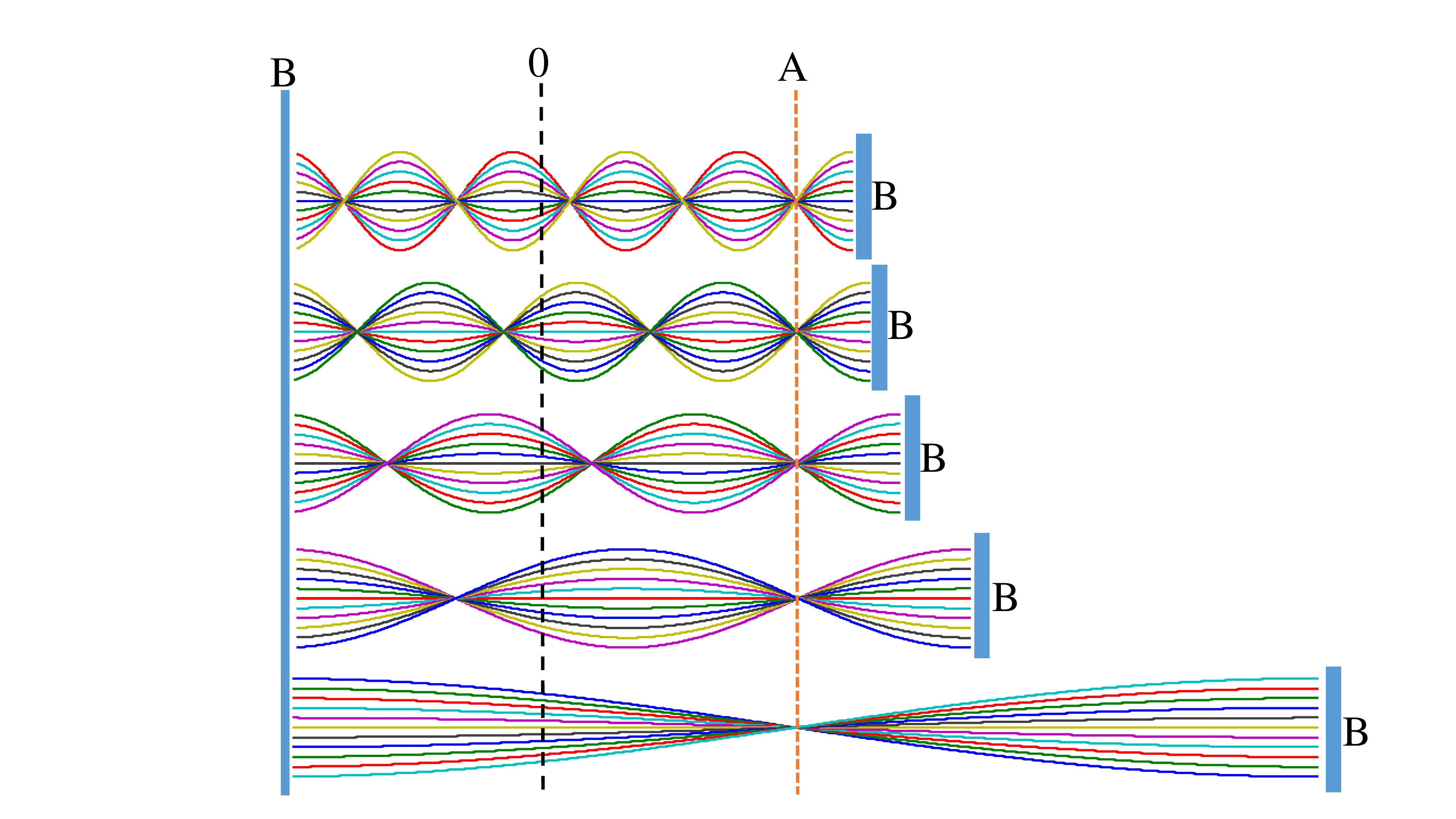}

\caption{(color online) A sketch of case (c) in Fig. \ref{FPsym1} clarified by replacing the only metallic mirror (dashed orange lines marked ``A'') by a phase conserving mirror (blue bars marked ``B''), retracted by a quarter wavelength, i.e. a frequency dependent cavity length.\label{FPsym3}}
\end{figure}

On the specific example of gap states created in a quasicrystalline Fibonacci chain, it is easy to understand the full meaning and importance of the notion of a $k$-dependent effective cavity length previously discussed. In fact, the use of a $k$-dependent cavity length as defined in (\ref{effect_length}) is required even in Fabry-Perot cavities with regular mirrors. In order to clarify this last statement, let us discuss two standard cases for a Fabry-Perot cavity of (geometrical) length $L$ depicted in Figs. \ref{FPsym1}a and \ref{FPsym1}b. The first is delimited by two dielectric  mirrors with a constant dielectric mismatch which conserves phase. The second contains two metallic mirrors, where each mirror contributes a reflected phase shift of $\pi$  which is frequency independent (up to the plasma frequency). The Fabry-Perot resonant frequencies for both cases are well predicted without the use of any phase shifts, by the standard formula $2L/\lambda_{m}\!=\! m$ where $m\!\in\!\mathbb{Z}$, with the lowest order nontrivial state at $\lambda_{1}=2L$. However, metallic and dielectric mirrors provide different boundary conditions forcing nodes and anti-nodes in the electric field envelope, respectively. Therefore, in the dielectric index mismatch phase conserving mirrors case (Fig. \ref{FPsym1}a), odd states are anti-symmetric and even states are symmetric (with respect to the mid-cavity coordinate) as expected. However, in the metallic mirrors case (Fig. \ref{FPsym1}a), the relation between spatial symmetry and the parity of $m$ is the other way around. Moreover, if we now consider a third non-standard Fabry-Perot cavity of cavity length $L$ with a dielectric mirror at one side and a metallic mirror on the other displayed in Fig. \ref{FPsym1}c, we see that resonant states are now asymmetric, and the lowest order state occurs at at $\lambda=4L$, unaccounted for by the usual formula $2L/\lambda_{m}\!=\! m\!\in\!\mathbb{Z}$. These deviations from the standard view are well understood when using the effective Fabry-Perot cavity length assigning an additional $\frac{\lambda}{4}$ virtual length for each metallic mirror following (\ref{effect_length}), and the resonant condition given by (\ref{FP_res3}). In the all dielectric cavity, the cavity phase shift is zero, i.e. ${\cal L}\left(\lambda\right)\equiv L$, and resonant states occur at $\lambda_{m}=\frac{2L}{m}$, with electric field anti-nodes at the boundaries, and with anti-symmetric (symmetric) odd (even) states with respect to the mid-cavity coordinate (Fig. \ref{FPsym1}a). In the all metal cavity case, we have ${\cal L}\left(\lambda\right)\equiv L+\frac{\lambda}{2}$, as each metallic mirror effectively extends the cavity by $\frac{\lambda}{4}$ (Fig. \ref{FPsym2}). The resonant states are retrieved by solving $\lambda_{m}=2{\cal L}(\lambda_{m})/m$ self consistently to arrive at $\lambda_{m}=\frac{2L}{m-1}$ which gives identical frequencies to the all dielectric case but for the opposite parity. This result, along with the fact that the effective and the geometrical cavity center coordinates coincide, fully explains the spatial properties of resonant states in an all metal Fabry-Perot cavity. In the hybrid cavity case, we have ${\cal L}\left(\lambda\right)\equiv L+\frac{\lambda}{4}$, as only one (metallic) mirror effectively extends the cavity by $\frac{\lambda}{4}$ (Fig. \ref{FPsym3}). Solving $\lambda_{m}=2{\cal L}(\lambda_{m})/m$ self consistently gives $\lambda_{m}=2L/(m-\frac{1}{2})$ yielding a different set of resonant frequencies than previous cases, including $\lambda_{1}=4L$. As the effective cavity center coordinate is shifted by $\frac{\lambda}{8}$ from the geometrical cavity center coordinate, symmetric and anti-symmetric states for the effective cavity appear asymmetric with respect to the geometrical cavity center coordinate.



\section{Discussion}

We have presented a method which allows to determine the spectral and spatial properties of gap states in photonic and electronic structures having a complex and gapped spectrum. This method is a generalisation of the standard Fabry-Perot calculation according to which relevant spectral frequencies result from a constructive interference condition. We have shown that this condition is equivalent to the integer winding of a properly defined phase characteristic of the Fabry-Perot cavity and which encapsulates its relevant details, e.g. its reflectance. We have presented a complete framework to determine systematically this cavity phase using scattering theory. We have recalled that the unitary scattering matrix contains all the relevant information about one or several scatterers in the form of a set of phases, precisely two for 1D systems studied here. One phase, the total phase shift, allows to determine in an elegant and practical way the change of spectral properties of large systems submitted to a given perturbation or a modification of boundary conditions. It is extensively used in a variety of problems such as the Casimir effects (static and dynamic) \cite{Matloob2001,Genet2003,Lambrecht2006}, cavity optomechanics \cite{Aspelmeyer2014}, surface physics, molecular or atomic (e.g. van der Waals) interactions or the amplification of spontaneous emission (Purcell effect) \cite{akkermans2013evgeni}. 

The second phase of the scattering matrix, independent of the total phase shift, is less ubiquitous. Nevertheless, we have shown that it allows to reformulate the Fabry-Perot constructive interference condition as a type of Levinson theorem, and also to properly account for existing symmetries and topological properties of the cavity. We have shown that in general, non trivial spectral and spatial properties of the cavity modes can be expressed in terms of those of the mirrors in the form of a generalised Fabry-Perot interference condition with a frequency-dependent cavity length. We have applied this approach to the study of topological properties of a 1D Fibonacci quasicrystal. Its generalisation to cavities bounded by mirrors with topological properties may find unexpected applications precisely in Casimir physics or optomechanics where topological features may allow to change possibly in a continuous way the nature of the interaction between the mirrors and thus the sign of the resulting Casimir force.

\subsubsection{Acknowledgments}

This paper is dedicated to the memory of Roger Maynard, a mentor and an outstanding physicist who was always prompt to get enthusiastic and encouraging about new ideas and new points of view. This work was supported by the Israel Science Foundation Grant No.924/09.

\section{References}

\bibliography{FPBIB}
\bibliographystyle{iopart-num}

\end{document}